\documentclass[12p]{amsart}

\usepackage[applemac]{inputenc}
\usepackage{color}

\usepackage{mathptmx}       
\usepackage{helvet}         
\usepackage{courier}        
\usepackage{type1cm}        
\usepackage{graphicx}        
\usepackage{multicol}        
\usepackage[bottom]{footmisc}

\usepackage{amsmath,amsfonts,amssymb}
\usepackage{mathtools}
\usepackage{bm} 

\usepackage{relsize,exscale}

\usepackage{url}

\newcommand{\ba}{\mathbf{a}}
\newcommand{\bd}{\mathbf{d}}

\newcommand{\bF}{\mathbf{f}}

\newcommand{\bh}{\mathbf{h}}
\newcommand{\bn}{\mathbf{n}}
\newcommand{\br}{\mathbf{r}}
\newcommand{\bs}{\mathbf{s}}
\newcommand{\bt}{\mathbf{t}}
\newcommand{\bu}{\mathbf{u}}
\newcommand{\bx}{\mathbf{x}}

\newcommand{\bB}{\mathbf{B}}
\newcommand{\bC}{\mathbf{C}}
\newcommand{\bD}{\mathbf{D}}
\newcommand{\bE}{\mathbf{E}}
\newcommand{\bH}{\mathbf{H}}
\newcommand{\bK}{\mathbf{K}}
\newcommand{\bL}{\mathbf{L}}
\newcommand{\bN}{\mathbf{N}}
\newcommand{\bR}{\mathbf{R}}
\newcommand{\bS}{\mathbf{S}}
\newcommand{\bX}{\mathbf{X}}
\newcommand{\bW}{\mathbf{W}}

\newcommand{\mB}{\mathcal{B}}
\newcommand{\mD}{\mathcal{D}}
\newcommand{\mF}{\mathcal{F}}

\newcommand{\mN}{\mathcal{N}}
\newcommand{\mR}{\mathcal{R}}

\newcommand{\mT}{\mathcal{T}}
\newcommand{\mU}{\mathcal{U}}
\newcommand{\mV}{\mathcal{V}}

\newcommand{\bepsilon}{\bm{\epsilon}}
\newcommand{\bkappa}{\bm{\kappa}}
\newcommand{\blambda}{\bm{\lambda}}
\newcommand{\bnabla}{\bm{\nabla}}
\newcommand{\bnu}{\bm{\nu}}
\newcommand{\brho}{\bm{\rho}}
\newcommand{\bsigma}{\bm{\sigma}}
\newcommand{\bSigma}{\bm{\Sigma}}
\newcommand{\bxi}{\bm{\xi}}

\newcommand{\asmby}{\mathsf{A}}

\title[]{A stochastic multi-scale approach for numerical modeling of complex materials -- Application to uniaxial cyclic response of concrete}

\date{}

\begin{document}

\maketitle

\renewcommand*{\thefootnote}{\fnsymbol{footnote}}

\begin{center}
Pierre~Jehel\textsuperscript{1,2}\footnote{Corresponding author: pierre.jehel[at]centralesupelec.fr} \\
\vspace{0.5cm}
$^1$ MSSMat, CNRS, CentraleSup\'elec, Universit\'e Paris-Saclay, Grande voie des Vignes, 92290 Ch\^atenay-Malabry, France\\
$^2$ Department of Civil Engineering and Engineering Mechanics, Columbia University, 630 SW Mudd, 500 West 120th Street, New York, NY, 10027, USA \\
\vspace{0.5cm}
\small{Manuscript form of the book chapter published online (Feb. 15, 2016):} \\
\small{\url{http://link.springer.com/book/10.1007/978-3-319-27996-1}} \\
\small{\copyright Springer International Publishing Switzerland 2016} \\
\small{A. Ibrahimbegovic (ed.), Computational Methods for Solids and Fluids} \\
\small{Computational Methods in Applied Sciences 41, chap. 6, pp. 123--160.} \\
\small{DOI: 10.1007/978-3-319-27996-1\_6}
\end{center}

\renewcommand*{\thefootnote}{\arabic{footnote}}

\begin{abstract}
In complex materials, numerous intertwined phenomena underlie the overall response at macroscale. These phenomena can pertain to different engineering fields (mechanical, chemical, electrical), occur at different scales, can appear as uncertain, and are nonlinear. Interacting with complex materials thus calls for developing nonlinear computational approaches where multi-scale techniques that grasp key phenomena at the relevant scale need to be mingled with stochastic methods accounting for uncertainties. In this chapter, we develop such a computational approach for modeling the mechanical response of a representative volume of concrete in uniaxial cyclic loading. A mesoscale is defined such that it represents an equivalent heterogeneous medium: nonlinear local response is modeled in the framework of Thermodynamics with Internal Variables; spatial variability of the local response is represented by correlated random vector fields generated with the Spectral Representation Method. Macroscale response is recovered through standard homogenization procedure from Micromechanics and shows salient features of the uniaxial cyclic response of concrete that are not explicitly modeled at mesoscale.
\end{abstract}

\section{Introduction}
\label{sec:1}
Widely-used materials in engineering practice such as polymer, composite, steel, concrete, are characterized by engineering parameters for design purposes, while these latter homogeneous macroscopic mechanical properties actually result from heterogeneous structures at lower scales. Material can be qualified as complex as their macroscopic behavior result from numerous multi-scale intertwined phenomena that have nonlinear and uncertain evolution throughout loading history. Modifications in the underlying structures of this category of materials can result in dramatic changes in mechanical behavior at the relevant macroscopic scale for engineering applications. Micro-cracks coalescence in the constitutive material of a structure challenges its capacity for meeting the performance level targeted during its design process. Alkali-aggregate reaction in concrete microscopic structure can lead to hazardous loss of bearing capacity in reinforced concrete structures. Accounting for phenomena at lower scales to reliably predict macroscopic response of heterogeneous structures is one of the challenges numerical multi-scale simulation techniques have been developed for over the past (\cite{BenDhia1998, LadLoiDur2001, Fey2003, IbrMar2003} among many others).

In continuum mechanics, explicitly accounting for relevant mechanisms and structures in heterogeneous scales underlying macroscopic scale provides the rationale for representing characteristic features of homogenized material behavior laws at macroscale that can then be used for engineering design. Micromechanics has been developed to extract macroscopic local continuum properties from microscopically heterogeneous media through the concept of Representative Volume Element (RVE). An RVE for a material point at macroscale is statistically representative of the microscopic structure in a neighborhood of this point~\cite{NemNasHor1993}. Also, Thermodynamics with Internal Variables provides a robust framework for modeling material response at macroscale according to a set of internal variables that carry information about the history of evolution mechanisms at lower scales without explicitly representing them~\cite{GerNguSuq1983, Maugin1999}. Other strategies to derive macroscopic mechanical properties of heterogeneous materials have been developed based on the introduction of a mesoscale, that is a scale that bridges the micro- and macroscales. In~\cite{OstSta2006}, heterogeneities are represented by random fields introduced at a mesoscale, which defines so-called Statistical Volume Elements that tends to become RVEs as mesoscale grows; effective properties at macroscale are retrieved according to two hierarchies of scale-dependent bounds obtained from either homogenous displacements or homogenous tensions applied on the boundary of the mesoscale. In~\cite{Ben-et-al2010}, a mesoscale is explicitly constructed for representing the macroscopic behavior of heterogeneous quasi-brittle materials. This mesoscale consists of a 3D finite element mesh composed of truss elements cut by inclusions. Truss element kinematics is enriched to account for discontinuities in the strain field due to the presence of inclusions along truss elements as well as discontinuities in the displacement field to account for possible cracks in the matrix, in the inclusions, or at their interface. With the improvement of computational ressources, stochastic homogenization of random heterogeneous media can now be achieved without introducing a mesoscale. In~\cite{Cot2013}, an efficient numerical strategy is presented to obtain effective tensors of random materials by coupling random micro-structures to tentative effective models within the Arlequin framework for model superposition~\cite{BenDhia1998}. In~\cite{Sav-et-al2014}, micro-structures composed of a medium with randomly distributed inclusions of random shapes are generated and their behaviors are simulated with the extended finite element method (XFEM); homogenized properties at macroscale are then derived through the computation of mean response using Monte Carlo simulations.

\begin{figure}[htb]
 \includegraphics[width=0.5\textwidth]{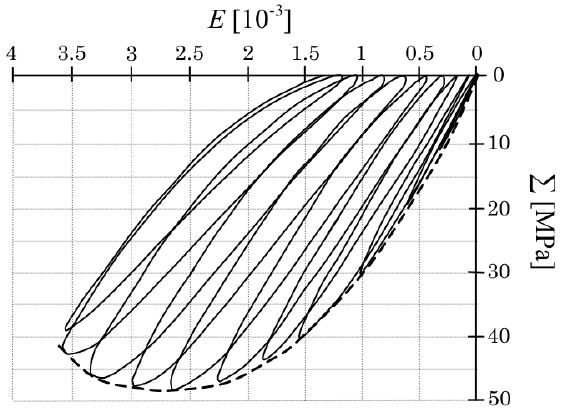}
 \caption{Strain-stress concrete experimental response in pseudo-static cyclic uniaxial compressive loading (adapted from~\cite{Ramtani1990}).}
 \label{fig:Ramtani}
\end{figure}

In the work presented in this chapter, we focus on the numerical representation of the homogenized one-dimensional response of a concrete specimen in cyclic compressive loading, as it can be observed in lab tests. Figure~\ref{fig:Ramtani} illustrates the main features of such an homogenized response: a backbone curve (dashed line) that is a nonlinear strain hardening phase ($0 \leq E \leq 2.7 \times 10^{-3}$) followed by a strain softening phase where strength degradation is observed; unloading-reloading cycles show that stiffness decreases while loading increases, hysteresis loops are generated. This typical response is observed at macroscale and results from numerous underlying mechanisms of physical or chemical nature at many different scales. For designing concrete structures, an equivalent homogeneous concrete model is sought, which has to represent concrete mechanical behavior in different loading conditions while accounting for mechanisms at lower scales~\cite{TorPijRey2012}. Heterogeneities can be observed in concrete at different scales: aggregates of different sizes are distributed in a cement paste; the so-called interfacial transition zone where the aggregates are bound to the cement paste plays a key role in the concrete mechanical properties~\cite{TriLib2014}; cement paste is composed of water, voids and of the products of the complete or partial hydration of the clinker particles, which generates a microscopic structure composed of numerous intertwined phases.

This chapter presents the basic ingredients of a stochastic multi-scale approach developed to represent the macroscopic compressive cyclic response of a concrete specimen while attempting not to sacrifice too much of the complexity of this material. To that aim, two scales are considered: the macroscale where an equivalent homogenous concrete model capable of representing the main features that are shown in figure~\ref{fig:Ramtani} is retrieved, and a mesoscale where heterogeneous local nonlinear response is assumed. Local response at mesoscale is modeled in the framework of Thermodynamics with Internal Variables and is seen as the homogenized response of mechanisms that occur at the micro- or nano- underlying scales. Spatial variability at mesoscale is introduced using stochastic vector fields. Homogenized macroscopic response is recovered using standard averaging method from micromechanics. 

The chapter is organized as follow. In the next section, the ingredients of the proposed stochastic multi-scale modeling are presented. First, the averaging method for computing the homogenized model response at macroscale is recalled. Then, the model of the mechanical local behavior of a material point at mesoscale is constructed. Finally, the Spectral Representation Method for generating stochastic vector fields that model heterogeneity at mesoscale is presented. In a third section, the numerical implementation of the approach in the framework of the finite element method is detailed. Before the conclusion, numerical applications are presented to demonstrate the capability of the proposed approach i) for yield homogeneous material behavior at macroscale without stochastic homogenization and ii) for representing salient features of macroscopic 1D concrete response in uniaxial cyclic compressive loading.

\section{Multi-scale stochastic approach for modeling concrete}
\label{sec:2}

\begin{figure}
 \includegraphics[width=1.0\textwidth]{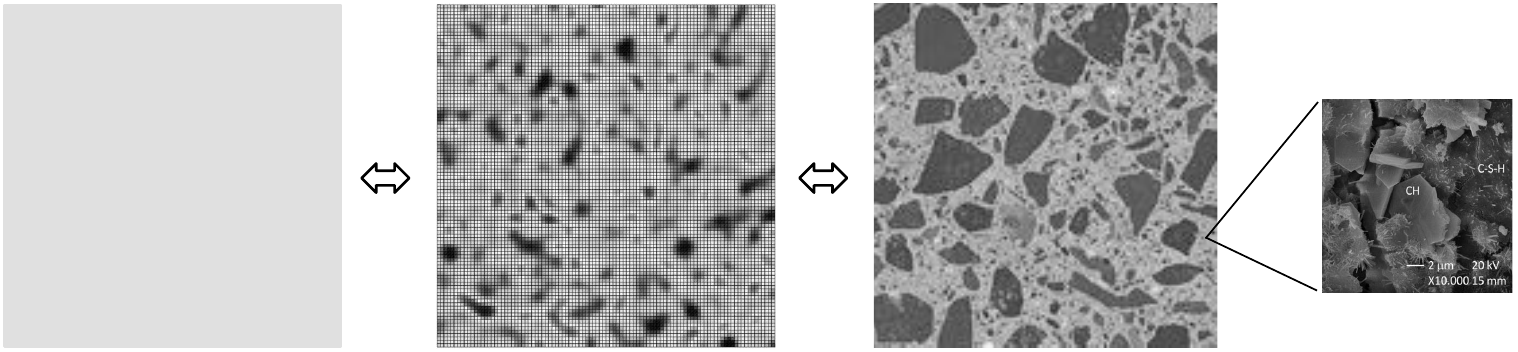}
 \caption{From left to right: equivalent homogeneous concrete (macroscale), equivalent heterogeneous concrete (E-mesoscale, 5~cm $\times$ 5~cm-square), actual heterogeneous concrete (A-mesoscale, 5~cm $\times$ 5~cm-square), and zoom on the underlying microstructure in the cement paste (20~$\mu$m $\times$ 20~$\mu$m-square observed through Scanning Electron Microscope, courtesy A.P.M. Trigo~\cite{TriLib2014}).}
 \label{fig:2scales}
\end{figure}

Figure~\ref{fig:2scales} presents the three following concepts, which further developments are based on:
\begin{itemize}
 \item \textit{Actual heterogeneous medium} (A-mesoscale): Concrete is made of aggregates distributed in a cement paste. Aggregates, cement and interface between both of them exhibit different mechanical responses. In the cement paste, micro- and nano-structures also exist.
 \item \textit{Equivalent heterogeneous medium} (E-mesoscale): The proposed approach does not consist in explicitly generating a multi-phase medium with random distribution of aggregates of random geometry in a cement paste with known mechanical behavior for each phase. The approach followed here consists in generating a random medium at each point of which the mechanical response obeys a prescribed behavior that has uncertain parameters and that is the homogenized response of mixtures of aggregates and cement where mechanisms at lower scales are also involved but not explicitly modeled.
 \item \textit{Equivalent homogeneous medium} (macroscale): Homogenization of E-mesoscale yields homogenized homogeneous concrete response. It will be shown in the numerical applications that one realization only of the random E-mesoscale can be sufficient to retrieve homogenized properties at macroscale.
\end{itemize}

\subsection{Homogenized material behavior at macroscale}

We consider a material elementary domain (ED) that occupies a spatial domain $\mR \subset \mathbb{R}^3$. The boundary $\partial\mR$ of the ED has outward normal $\bn$, tension $\bar{\bt}$ can be imposed on the part of the boundary $\partial_{\sigma}\mR$ while displacement $\bar{\bu}$ can be imposed on $\partial_u\mR$, where $\partial_{\sigma}\mR \cup \partial_u\mR = \partial\mR$ and $\partial_{\sigma}\mR \cap \partial_u\mR = \emptyset$. There are no external forces other than $\bar{\bt}$ applied on the ED and no dynamic effects are considered either. Then, the displacement vector field $\bu$, and the strain and stress tensor fields, $\bepsilon$ and $\bsigma$, satisfy at any pseudo-time $t \in [0, \ T]$:
\begin{align} \label{eq:BVPs}
 & \mathbf{div} \ \bsigma(\bx,t) = \mathbf{0} & & \forall \bx \in \mR \nonumber \\
 & \bepsilon(\bx,t) = \mathrm{sym}\left[\bnabla(\bu(\bx,t))\right] & & \forall \bx \in \mR \nonumber \\
 & \bsigma(\bx,t) = \hat{\bsigma}(\bepsilon(\bx,t)) & & \forall \bx \in \mR \\
 & \bsigma(\bx,t) \cdot \bn(\bx) = \bar{\bt}(\bx,t) & & \forall \bx \in \partial_{\sigma}\mR \nonumber \\
 & \bu(\bx,t) = \bar{\bu}(\bx,t) & & \forall \bx \in \partial_u\mR \nonumber
\end{align}
$\mathrm{sym}\left[\bnabla(\cdot)\right] := \frac{1}{2}\left(\bnabla(\cdot)+\bnabla^T(\cdot)\right)$ is the symmetric part of the gradient tensor $\bnabla(\cdot)$, the superscript $(\cdot)^T$ denoting the transpose operation. In the set of equations above, small strains are assumed and behavior law $\hat{\bsigma}(\bepsilon)$ can be nonlinear.

We classically assume that any macroscopic quantity $Q$ is connected to its E-mesoscopic counterpart $q$ through domain averaging over the ED:
\begin{equation}
 Q(\bX) := \langle q \rangle(\bX) = \frac{1}{\vert \mathcal{R} \vert} \int_{\mathcal{R}} q(\bx; \bX) d\bx
\end{equation}
$\vert \mathcal{R} \vert = \int_{\mR} d\bx$ is the measure of the spatial domain occupied by the ED centered at material point $\bX$ of the macroscale, and $\bx$ denotes a material point of the E-mesoscale.

In all what follows, we will assume linear displacements imposed all over the boundary of $\mR$:
\begin{equation}\label{eq:hom-BCs}
 \bu(\bx,t) = \bE^{\prime}(\bX,t) \cdot \bx \quad ; \quad \forall \bx \in \partial\mR \ , \ \forall t \in [0,\,T]
\end{equation}
Hence, $\partial_u\mR = \partial\mR$ and $\partial_{\sigma}\mR = \emptyset$. With this assumption, it can be shown (see e.g.~\cite[Chap.~1]{NemNasHor1993} or~\cite{Zaoui2002}) that:
\begin{equation}\label{eq:eps-ave}
 \bE^{\prime}(\bX,t) = \bE(\bX,t) := \langle \bepsilon(\bx,t) \rangle \quad ; \quad \bx \in \mR(\bX)
\end{equation}
and also, because it is assumed there is no external forces applied on $\mR(\bX)$:
\begin{equation} \label{eq:Sig-from-t}
  \bSigma(\bX,t) = \frac{1}{\vert \mR \vert} \int_{\partial\mR} \mathrm{sym}\left[\bt(\bx,t) \otimes \bx\right] \ d\partial\mR
 \end{equation}
where $\bt(\bx,t) := \bsigma(\bx,t) \cdot \bn(\bx)$ are the tension forces developed over $\partial_u\mR$.

Note that other boundary conditions could be considered. In any case, it is in general not possible to derive the strain or stress fields at E-mesoscale from the macroscopic quantities, and consequently simplifying assumptions as in~\eqref{eq:hom-BCs} are made. Whether it is displacements or forces that are imposed on $\mR$, and whether these latter conditions are linear or periodic, this can influence the homogenized macroscopic response of the ED. However, this is out of the scope of this work where the consequences of assuming linear displacements imposed on $\partial\mR$ will not be discussed.

With the boundary conditions~\eqref{eq:hom-BCs} applied on $\mR$, Hill's lemma can be proved:
\begin{equation} 
 \langle \bsigma : \bepsilon \rangle = \langle \bsigma \rangle : \langle \bepsilon \rangle := \bSigma : \bE
\end{equation}
which means that the medium recovered at macroscale through homogenization is energetically equivalent to the heterogeneous medium considered at E-mesoscale. 

The possibly nonlinear material response at mesoscale is expressed as:
\begin{equation} \label{eq:mes-tan-mod}
 \dot{\bsigma} = \blambda : \dot{\bepsilon}
\end{equation}
where $\blambda$ is the tangent modulus at E-mesoscale and the superimposed dot denotes partial derivative with respect to pseudo-time. Thanks to Hill's lemma, we then have the following two equivalent definitions for the tangent modulus $\bL$ at the homogenized macroscale:
\begin{equation} \label{eq:mac-tan-mod}
 \langle \bepsilon : \blambda : \dot{\bepsilon} \rangle = \bE : \bL : \dot{\bE} \qquad \Leftrightarrow \qquad \dot{\bSigma} = \bL : \dot{\bE}
\end{equation}

\subsection{Material behavior law at E-mesoscale}
\label{sec:MatModMes}

\begin{figure}[htb]
 \includegraphics[width=1.0\textwidth]{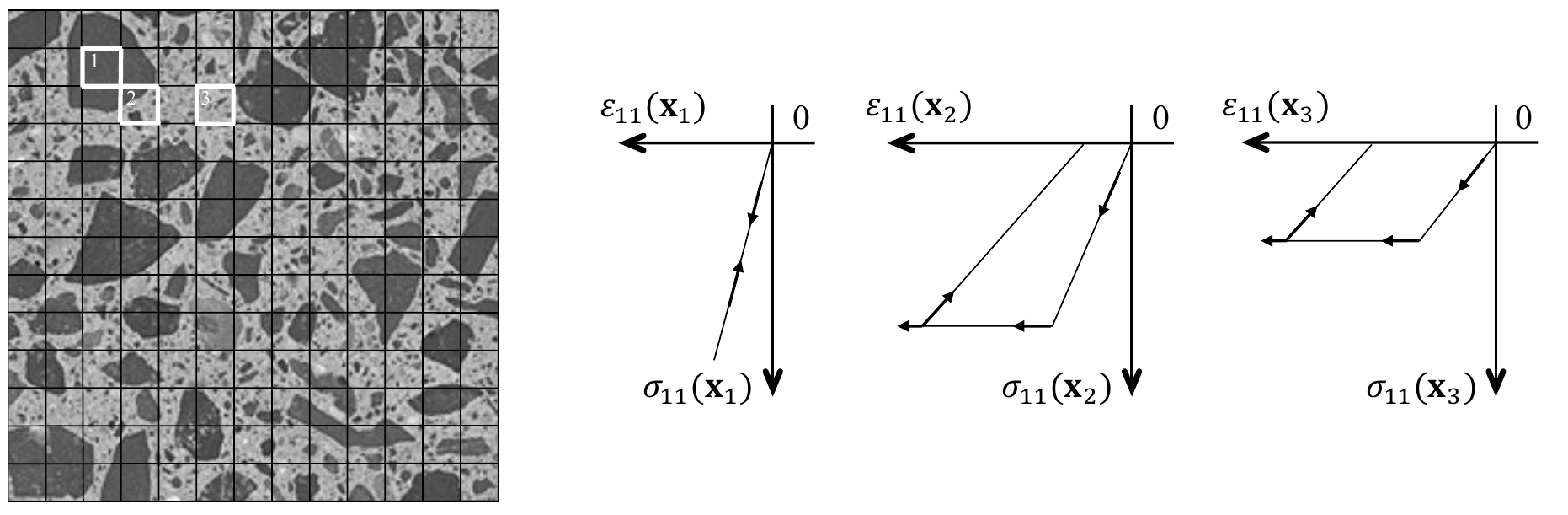}
 \caption{Each point at E-mesoscale has a different behavior due to the heterogenous structure of concrete. Behavior laws at E-mesoscale are homogenized responses of aggregate-cement paste mixtures with also heterogenous microstructures in the cement paste. At point $\bx_1$ there is an aggregate solely; at point $\bx_2$ there is mixture of large and small aggregates in the cement paste; at point $\bx_3$ there is mainly cement with some small aggregates.}
 \label{fig:dam-pla-r}
\end{figure}

We assume a coupled damage-plasticity model to be suitable for representing material response at any material point $\bx$ at E-mesoscale (see figure~\ref{fig:2scales}). This choice is motivated by the fact that concrete A-mesoscale is composed of both a ductile cement matrix that can be represented by a plastic model, and brittle aggregates that are confined in the cement paste and whose compressive response can be more realistically represented by a damage model. Hereafter, we develop a model in a way that allows for explicitly controlling the coupling of damage and plasticity. Indeed, as illustrated in figure~\ref{fig:dam-pla-r}, the response of material points at mesoscale can be either better represented by a damage model alone, or a plasticity model alone, or by the appropriate coupling of both models. This is developed in the framework of thermodynamics with internal variables~\cite{GerNguSuq1983, Maugin1999} where the internal variables carry the history of irreversible mechanisms occurring in the material at lower (micro- and nano-) scales.

\subsubsection{Basic ingredients}

The three basic ingredients for developing this model of local behavior law at E-mesoscale are as follows:
\begin{itemize}
 \item Total deformation $\bepsilon$ is split into damage ($\bepsilon^d$) and plastic ($\bepsilon^p$) parts:
\begin{equation}\label{eq:def-split}
 \bepsilon := \bepsilon^d + \bepsilon^p
\end{equation}

 \item Stored energy function is defined as:
\begin{equation}
 \psi(\bepsilon, \bD, \bepsilon^p) := \bsigma : \bepsilon^d -  \frac{1}{2} \bsigma : \bD : \bsigma
\end{equation}
with $\bD$ the fourth-order damage compliance tensor. $\bD$ and $\bepsilon^p$ are the internal variables that drive the evolution of the material. Also, denoting by $\bC$ the elasticity tensor, we set initially, as the material is undamaged, $\bD^{-1} = \bC$. The elements of $\bC$ are parameters of the model.

 \item A criterium function is introduced as:
\begin{equation}\label{eq:cri-fun}
 \phi(\bsigma) := h(\bsigma) - \sigma_y \leq 0
\end{equation}
It defines the limit states between the states where there is no evolution of the internal variables ($\phi < 0$) and those where there is evolution ($\phi = 0$). The so-called yield stress $\sigma_y > 0$ is a scalar parameter.
\end{itemize}

More general models coupling damage and plasticity with hardening or softening could be defined. Then, other internal variables would be introduced (see e.g.~\cite{MarIbr2006, IbrJeh2008, Jeh-et-al2010}).

\subsubsection{Material dissipation and state equation}

Then, the material dissipation reads:
\begin{eqnarray} \label{eq:mat-dis}
 \mD &:=& \bsigma : \dot{\bepsilon} - \dot{\psi} \geq 0 \nonumber \\
 &=& \dot{\bsigma} : \left( \bD : \bsigma - \bepsilon^d \right) + \frac{1}{2} \bsigma : \dot{\bD} : \bsigma + \bsigma : \dot{\bepsilon}^p \geq 0
\end{eqnarray}
$\mD$ should be non-negative to comply with the principle of thermodynamics. In case there is no evolution of the internal variables, that is for loading steps that do not generate any change of state in the material, there is no evolution of the internal variables: $\dot{\bD} = \dot{\bepsilon}^p =\mathbf{0}$ and the process is assumed to be non-dissipative, that is $\mD$ is null. According to equation~\eqref{eq:mat-dis}, it then comes the state equation:
\begin{equation}\label{eq:sta-equ}
 \bepsilon^d := \bD : \bsigma
\end{equation}
Equation~\eqref{eq:sta-equ} is to this damage model what the more classical constitutive relation $\bsigma := \bC : \bepsilon^e$ is to linear elasticity model.
 
Introducing this latter state equation into equation~\eqref{eq:mat-dis}, we can rewrite:
\begin{equation}
 \mD = \frac{1}{2} \bsigma : \dot{\bD} : \bsigma + \bsigma : \dot{\bepsilon}^p \geq 0
\end{equation}
from where we define $\mD^d := \frac{1}{2} \bsigma : \dot{\bD} : \bsigma \geq 0$ and $\mD^p := \bsigma : \dot{\bepsilon}^p \geq 0$.

\subsubsection{Evolution of the internal variables}

Following what has been done to derive the equations of mechanical models with plasticity solely~\cite{Hill1950}, the evolution of the internal variables is obtained appealing to the principle of maximum dissipation. Accordingly, among all the admissible stresses, that is $\bsigma$ such that $\phi(\bsigma) \leq 0$, it is those that maximize the material dissipation $\mD$ that have to be retained. This can be cast into a minimization problem with constraint $\phi \leq 0$~\cite{Lubliner1984}. Lagrange multiplier method can be used to solve it with the so-called Lagrangian reading:
\begin{eqnarray}
 \mathcal{L}(\bsigma,\dot{\gamma}) &:=& -\mD + \dot{\gamma} \, \phi \nonumber \\
 &=& ( -\mD^d + \dot{\gamma}^d \, \phi ) + ( -\mD^p + \dot{\gamma}^p \, \phi )
\end{eqnarray}
Here, we have split the total Lagrange multiplier $\dot{\gamma} \geq 0$ so that $\dot{\gamma} = \dot{\gamma}^d + \dot{\gamma}^p$ with two Lagrange multipliers defined as $\dot{\gamma}^d := r \, \dot{\gamma}$ and $\dot{\gamma}^p := (1-r) \, \dot{\gamma}$ where $r$ is to be taken in the range $[0, \ 1]$. $r$ is a damage-plasticity coupling parameter: if $r=0$, $\dot{\gamma}^d=0$ and there is plasticity evolution only; if $r=1$, only damage evolves in the material; and for any other $r$ in-between, there is coupled evolution of both damage and plasticity.

In turn, the Lagrangian is also split into damage and plasticity parts:
\begin{equation}
 \mathcal{L}^d(\bsigma,\dot{\gamma}^d) := -\frac{1}{2} \bsigma : \dot{\bD} : \bsigma + \dot{\gamma}^d \, \phi \qquad ; \qquad \mathcal{L}^p(\bsigma,\dot{\gamma}^p) := -\bsigma : \dot{\bepsilon}^p + \dot{\gamma}^p \, \phi
\end{equation}
Both parts have to be minimized to ensure the total Lagrangian is minimum. The Kuhn-Tucker optimality conditions associated to these minimization problems result in:
\begin{equation}
 \frac{\partial \mathcal{L}^{d,p}}{\partial \bsigma} = \mathbf{0} \qquad \mathrm{and} \qquad \frac{\partial \mathcal{L}^{d,p}}{\partial \dot{\gamma}^{d,p}} = 0
\end{equation}
Setting $\bnu := \partial\phi \slash \partial\bsigma$, this leads to the following equations of evolution of the internal variables:
\begin{align}
 \dot{\bD} : \bsigma & = \dot{\gamma}^d \, \bnu := r \, \dot{\gamma} \, \bnu \label{eq:D-evo} \\
 \dot{\bepsilon}^p  & = \dot{\gamma}^p \, \bnu := (1-r) \, \dot{\gamma} \, \bnu \label{eq:epP-evo}
\end{align}
Besides, this minimizing problem also yields the following so-called loading/unloading conditions:
\begin{equation} \label{eq:loa-unl}
 \dot{\gamma}^{d,p} \geq 0 \ ; \ \phi \leq 0 \ ; \ \dot{\gamma}^{d,p} \, \phi = 0
\end{equation}

\subsubsection{Damage and plasticity multipliers}

In the case $\dot{\gamma}^d > 0$ or $\dot{\gamma}^p > 0$, there is damage or plasticity evolution and, according to~\eqref{eq:loa-unl}, $\phi(\bsigma)$ as to remain null during the process so that the stresses remain admissible. We thus have the consistency condition $\dot{\phi} = 0$ that can be rewritten as:
\begin{equation}\label{eq:con-con-1}
 \frac{\partial \phi}{\partial\bsigma} : \frac{\partial\bsigma}{\partial t} = \bnu : \dot{\bsigma} = 0
\end{equation}
Remarking from~\eqref{eq:sta-equ} that $\dot{\bepsilon}^d = \dot{\bD} : \bsigma + \bD : \dot{\bsigma}$ and using equations~\eqref{eq:def-split}, \eqref{eq:D-evo} and~\eqref{eq:epP-evo}, we have:
\begin{equation} \label{eq:dot-sig}
 \bD : \dot{\bsigma} = \dot{\bepsilon} - \dot{\gamma} \, \bnu
\end{equation}
Then, assuming $\bD \neq \mathbf{0}$, the consistency condition~\eqref{eq:con-con-1} is satisfied when $\dot{\gamma} > 0$ if: 
\begin{equation} \label{eq:gam-evo}
 \dot{\gamma} \, \bnu = \dot{\bepsilon}
\end{equation}
Or, with the damage and plasticity multipliers $\dot{\gamma}^d > 0$ and $\dot{\gamma}^p > 0$:
\begin{equation} \label{eq:gam-evo-pd}
 \dot{\gamma}^d \, \bnu = r \, \dot{\bepsilon} \qquad \mathrm{and} \qquad \dot{\gamma}^p \, \bnu = (1-r) \, \dot{\bepsilon}
\end{equation}

\subsubsection{Tangent modulus}
\label{sec:TanMod}

The tangent modulus at mesoscale $\blambda$ is a fourth-order tensor that has been defined in~\eqref{eq:mes-tan-mod} such that $\dot{\bsigma} = \blambda : \dot{\bepsilon}$. Assuming that $\bD^{-1}$, the inverse of $\bD$, exists ($\bD^{-1}:\bD=\mathbf{I}$, where $\mathbf{I}$ is the identity fourth-order tensor), and reminding equations~\eqref{eq:dot-sig} and~\eqref{eq:gam-evo}, we have:
\begin{equation} \label{eq:tan-mod}
 \blambda = \left\{
 \begin{array}{ll}
  \bD^{-1} \quad & \textrm{if} \quad \dot{\gamma} = 0 \quad (\phi(\bsigma) < 0) \\
  \mathbf{0} \quad &  \textrm{if} \quad \dot{\gamma} > 0 \quad (\phi(\bsigma) = 0 \ ; \ \dot{\phi}(\bsigma) = 0)
 \end{array} \right.
\end{equation}

\bigskip

\begin{figure}[htb]
 \includegraphics[width=.4\textwidth]{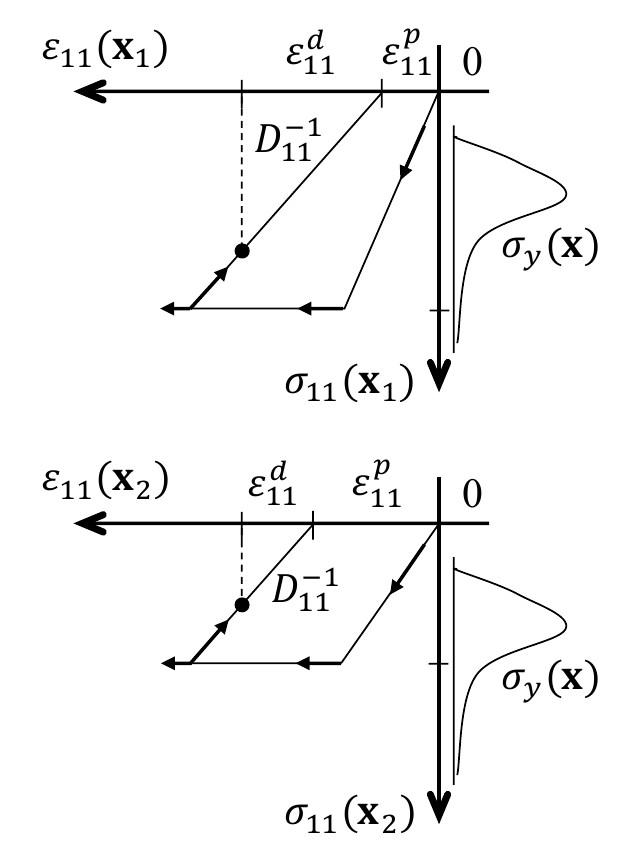}
 \caption{Example of the model response at two different material points of the E-mesoscale. Spatial variability is explicitly illustrated on the figure. Initial stiffness is determined by a spatially variable elastic modulus ($D^{-1}_{11}=C_{11}$); yielding threshold $\sigma_y$ fluctuates over $\mR$; how fast damage evolves comparing to plasticity is governed by the spatially variable coupling parameter $r$.}
 \label{fig:pnt-law}
\end{figure}

To sum up, the proposed material model at E-mesoscale is based on a set of internal variables that consists of the damage compliance tensor $\bD$ and the plastic deformation tensor $\bepsilon^p$. Besides, the model is parameterized by the elasticity tensor $\bC$, the stress threshold $\sigma_y$ above which damage or plastic evolution occurs, and the damage-plasticity coupling coefficient $r$: if $r=1$, there is no plastic evolution and the material can only damage, while if $r=0$, there is no damage evolution and the material is perfectly plastic. Figure~\ref{fig:pnt-law} shows material constitutive behavior at two different material points $\bx_1$ and $\bx_2$ of the E-mesoscale where the parameters take different values: parameters, and consequently local response, vary over the domain $\mR$ due to heterogeneities at E-mesoscale.

\subsection{Stochastic modeling of heterogeneous E-mesoscale}

Spatial variability at E-mesoscale of a set of $m$ parameters $\ba$ over $\mR$ is conveyed through stochastic modeling: it is assumed that the fluctuations of correlated stochastic fields can describe the actual material heterogeneous meso-structure (A-mesoscale). Thus, we introduce the probability space $(\Theta, \mathfrak{S}, P)$ where $\Theta$ is the sample space containing all the possible outcomes $\theta$ from the observation of the random phenomenon that is studied; $\mathfrak{S}$ is the $\sigma$-algebra associated with $\Theta$; $P$ is a probability measure. A real parameter $a \in \ba$ taking values in $\mV_a$ is then considered as the realization of a random variable $\mathfrak{a}(\theta)$: $\Theta \rightarrow \mV_a$. A random variable can be completely defined by its cumulative distribution function: $\mF_{\mathfrak{a}}(a) = \Pr[\mathfrak{a}(\theta) \leq a] = \int_{\{\theta \vert \mathfrak{a}(\theta) \leq a \}} P(\theta)$ or, when a probability density function (PDF) $p_{\mathfrak{a}}(a)$ exists: $\mF_{\mathfrak{a}}(a) = \int_{ \{ s \in \mV_a \vert s \leq a \} } p_{\mathfrak{a}}(s) \ ds$. 

Before we go on with the definition of stochastic fields, we recall some basic definitions for random variables. The mean $\mu_{\mathfrak{a}}$ and the variance $s^2_{\mathfrak{a}}$ of a random variable $\mathfrak{a}$ are defined as:
\begin{align}
 & \mu_{\mathfrak{a}} := \mathbb{E}[\mathfrak{a}] = \int_{-\infty}^{+\infty} a \, p_{\mathfrak{a}}(a) \, da  \\
 & s^2_{\mathfrak{a}} := \mathbb{E}[(\mathfrak{a} - \mu_{\mathfrak{a}})^2]
\end{align}
where $\mathbb{E}[\cdot]$ is the so-called mathematical expectation and $p_{\mathfrak{a}}(a) = 0$, $\forall a \in \mathbb{R} \setminus \mV_a$. Also, $\mathfrak{a}$ and $\mathfrak{b}$ being two random variables, the  covariance is defined as:
\begin{align} \label{eq:def-cov}
 Cov_{\mathfrak{a} \mathfrak{b}} & := \mathbb{E}[(\mathfrak{a} - \mu_{\mathfrak{a}}) (\mathfrak{b} - \mu_{\mathfrak{b}})] \nonumber \\
 & := \int_{-\infty}^{+\infty} \int_{-\infty}^{+\infty} (\mathfrak{a} - \mu_{\mathfrak{a}}) (\mathfrak{b} - \mu_{\mathfrak{b}}) \, p_{\mathfrak{a}\mathfrak{b}}(a,b) \, da \, db
\end{align}
where $p_{\mathfrak{a}\mathfrak{b}}$ is the joint PDF of $\mathfrak{a}$ and $\mathfrak{b}$, with $p_{\mathfrak{a}\mathfrak{b}}(a,b) = 0$ $\forall (a,b) \in \mathbb{R}^2 \setminus \mV_a \times \mV_b$. We also introduce the correlation, which is defined as:
\begin{equation}\label{eq:Rdef0}
 R_{\mathfrak{a} \mathfrak{b}} := \mathbb{E}[\mathfrak{a} \, \mathfrak{b}] = Cov_{\mathfrak{a} \mathfrak{b}} + \mu_{\mathfrak{a}} \, \mu_{\mathfrak{b}}
\end{equation}
And finally the following correlation coefficient will also be used later on:
\begin{equation} \label{eq:def-cor-coeff}
 \rho_{\mathfrak{a} \mathfrak{b}} := \frac{Cov_{\mathfrak{a} \mathfrak{b}}}{s_{\mathfrak{a}} \, s_{\mathfrak{b}}} \ \in [-1, \ 1]
\end{equation}

\subsubsection{Random vector fields for modelind heterogeneous meso-structure}

It is assumed that the heterogeneity of the parameters $\bC$, $\sigma_y$ and $r$ of the model developed above for representing material response at E-mesoscale over a concrete elementary domain (ED) can be represented as the realization of a random vector field. A random vector $\bm{a}(\theta)$ is a vector of random variables. Let $\mB \subset \mathbb{R}^d$ be a spatial domain of dimension $d$; this can be a volume ($d=3$), an area ($d=2$) or a length ($d=1$). A random vector field $\bm{g}(\bx; \theta)$ over $\mB$ is a collection of random vectors indexed by the position $\bx = (x_1,...,x_d)^T \in \mB$. For any fixed $\bx \in \mB$, any component $\mathfrak{g}_j(\bx)$ of $\bm{g}$, $j \in [1,...,m]$, is a random variable. In the case of random vector fields, we have the following definitions of the mean, the auto-correlation and cross-correlation functions respectively:
\begin{align}
 \mu^{\bm{g}}_j(\bx) & := \mathbb{E}[\mathfrak{g}_j(\bx)] & \quad & j \in [1,..,m] \\
 R^{\bm{g}}_{j j} (\bx,\bxi)  & := \mathbb{E}[\mathfrak{g}_j(\bx) \, \mathfrak{g}_j(\bx+\bxi)] & \quad & j \in [1,..,m] \\
 R^{\bm{g}}_{j k} (\bx,\bxi) & := \mathbb{E}[\mathfrak{g}_j(\bx) \, \mathfrak{g}_k(\bx+\bxi)] & \quad & j \in [1,..,m] \ , \ k \in [1,..,m] \ , \ j \neq k \label{eq:Rdef}
\end{align}
where $\bxi = (\xi_1,...,\xi_d)^T$ is the separation distance vector between two points of $\mB$. 

To fully characterize random vector fields, we need the marginal and joint PDFs of all possible combinations of random variables $\mathfrak{g}_j(\bx)$. From a practical point of view, this implies that many concrete ED have to be considered, that for each of them the parameters of interest have to be identified at many points $\bx$ all over $\mB$, so that these PDFs can be empirically constructed. If gathering such a huge amount of information was needed, the usefulness of using random vector field for modeling heterogeneity in concrete structure at mesoscale would be questionable. Therefore, we will make assumptions on the structure of the random vector field that would justify the efficiency of the proposed approach.

E-mesoscale construction will rely on the two following assumptions:
\begin{enumerate}
 \item Random fields will be generated as Gaussian. This means that random field $\mathfrak{g}_j(\bx;\theta)$ is fully characterized by both its mean function $\mu^{\bm{g}}_j(\bx)$ and auto-correlation function $R^{\bm{g}}_{j j} (\bx,\bxi)$. Nevertheless, non-Gaussian random field can then be obtained through nonlinear translation of Gaussian field, which will be discussed in section~\ref{sec:tra-RF}.
 \item Random fields are jointly homogeneous. This means that their mean function is independent of the position $\bx$ and that auto- and cross-correlation functions depend on the separation distance only:
\begin{align}
 \mu^{\bm{g}}_j(\bx) & := \mu^{\bm{g}}_j & \quad & j \in [1,..,m] \\
 R^{\bm{g}}_{j k} (\bx,\bxi) & := R^{\bm{g}}_{j k}(\bxi) & \quad & j \in [1,..,m] \ , \ k \in [1,..,m]
\end{align}
Note that efficient techniques can be used to account for heterogeneity in the random field (see e.g.~\cite{PapSoiPap2013}).
\end{enumerate}

Also, we will consider the possible ergodicity of the generated random fields in mean and correlation functions. One realization of such an ergodic random vector field contains all the statistical information needed to retrieve the first two moments: means and correlation functions can be computed as spatial averages.

\subsubsection{Spectral representation of homogeneous Gaussian random vector fields}

We present here the Spectral Representation Method for generating standard homogenous Gaussian fields~\cite{ShiDeo1991, ShiDeo1996, Deodatis1996a, Deodatis1996b, PopDeoPre1998}. Note that considering the Gaussian random fields to be standard, that is with zero mean and unit variance, does not introduce any loss of generality because non-standard Gaussian fields can always be retrieved through linear transformation.

The basic ingredient is the definition of a target correlation matrix, which can be built from experimental observations for instance:
\begin{equation}
 \bR^0(\bxi) =
 \begin{pmatrix}
  R^0_{11}(\bxi) & \cdots & R^0_{1m}(\bxi) \\
  \vdots & \ddots & \vdots \\
  R^0_{m1}(\bxi) & \cdots & R^0_{mm}(\bxi)
 \end{pmatrix}
\end{equation}
Superscript $^0$ has been added to highlight that these functions are target correlations, which should be retrieved in the statistical analysis of the generated random fields.

According to Wiener-Khinchin theorem, power spectral density functions $S^0_{jj}$, $j\in [1,...,m]$, and cross-spectral density functions $S^0_{jk}$, $(j,k) \in [1,...,m]^2, j \neq k$, are the Fourier transform of the corresponding correlation functions:
\begin{align}
 & S^0_{jk}(\bkappa) = \frac{1}{(2\pi)^d} \int_{-\infty}^{+\infty} \cdots \int_{-\infty}^{+\infty} R^0_{jk}(\bxi) \ e^{-i \bkappa \cdot \bxi} \ d\xi_1 \cdots d\xi_d \\
 & R^0_{jk}(\bxi) = \int_{-\infty}^{+\infty} \cdots \int_{-\infty}^{+\infty} S^0_{jk}(\bkappa) \ e^{i \bkappa \cdot \bxi} \ d\kappa_1 \cdots d\kappa_d
\end{align}
where $\bkappa = (\kappa_1,...,\kappa_d)^T$ is the wave number vector, $\bkappa \cdot \bxi$ is the scalar product of the two vectors $\bkappa$ and $\bxi$, and $i$ is the imaginary unit. Power spectral density functions are by definition real functions of $\bkappa$ while cross-spectral functions can be complex functions of $\bkappa$. It has been shown in~\cite{Shinozuka1987} that matrix $\bS(\bkappa)$ is Hermitian and semidefinite positive, which implies that it can be decomposed using Cholesky's method as:

\begin{equation}
 \bS(\bkappa) = \bH(\bkappa) \, \bH^{\star T}(\bkappa)
\end{equation}
where $(\cdot)^{\star}$ denotes the complex conjugate and $\bH(\bkappa)$ is a lower triangular matrix. The diagonal elements $H_{jj}$ are real and non-negative functions of $\bkappa$ while the off-diagonal elements can be complex:
\begin{equation}
 H_{jk}(\bkappa) = \vert H_{jk} \vert \ e^{i \, \varphi_{jk}(\bkappa)} \quad ; \quad j \in [1,...,m] \ ; \ k \in [1,...,m] \ ; \ j > k
\end{equation}

Then, the $j$\textsuperscript{th} component of a realization of a $m$V-$d$D homogeneous standard Gaussian stochastic vector field $\bm{g}(\bx;\theta)$ with cross-spectral density matrix $\bS(\bkappa)$ reads:
\begin{align}\label{eq:gVFcomp}
 g_j(\bx;\theta) = & 2 \sqrt{\Delta \kappa_1 \cdots \Delta\kappa_d} \ \sum_{l=1}^m \sum_{n_1=0}^{N_1-1} \cdots \sum_{n_d=0}^{N_d-1} \sum_{\alpha=1}^{2^{d-1}} \ \vert H_{jl}(\bkappa^{\alpha}_{n_1...n_d}) \vert \nonumber \\
 & \times \cos \left( \bkappa^{\alpha}_{n_1...n_d} \cdot \bx - \varphi_{jl}(\bkappa^{\alpha}_{n_1...n_d}) + \Phi_{l,n_1...n_d}^{\alpha}(\theta) \right)
\end{align}
for $j \in [1,...,m]$ and $N_1 \rightarrow +\infty$,...,  $N_d \rightarrow +\infty$. In equation~\eqref{eq:gVFcomp}, the following notation has been introduced:
\begin{equation}\label{eq:Dk}
 \bkappa^{\alpha}_{n_1...n_d} := \left( n_1 \, \Delta\kappa_1 \quad I_2^{\alpha} n_2 \, \Delta\kappa_2 \quad \ldots \quad I_d^{\alpha} n_d \, \Delta\kappa_d \right)^T
\end{equation}
Wave numbers increments are defined as:
\begin{equation}\label{eq:Delk}
 \Delta \kappa_i := \frac{\kappa_{u \, i}}{N_i} \quad , \quad i \in [1,...,d]
\end{equation}
where $\kappa_{u \, i}$'s are so-called cut-off wave numbers such that $\bS(\bkappa)$ can be assumed to be null for any $\kappa_i \geq \kappa_{u \, i}$. Also, $(I_1^{\alpha},I_2^{\alpha}, ..., I_d^{\alpha})$ are the $\alpha$ different vectors composed of $+1$'s and $-1$'s where $I_1^{\alpha}=1$ for all $\alpha$. For instance, for $d=3$, there are the following $\alpha = 2^{3-1}=4$ different such arrangements: $(1, 1,1)$, $(1,1,-1)$, $(1,-1,1)$ and $(1,-1,-1)$, so that $\bkappa^1_{n_1 n_2 n_3} = (n_1 \, \Delta\kappa_1, n_2 \, \Delta\kappa_2, n_3 \, \Delta\kappa_3)^T$, $\bkappa^2_{n_1 n_2 n_3} = (n_1 \, \Delta\kappa_1, n_2 \, \Delta\kappa_2, - n_3 \, \Delta\kappa_3)^T$, $\bkappa^3_{n_1 n_2 n_3} = (n_1 \, \Delta\kappa_1, - n_2 \, \Delta\kappa_2, n_3 \, \Delta\kappa_3)^T$ and $\bkappa^4_{n_1 n_2 n_3} = (n_1 \, \Delta\kappa_1, - n_2 \, \Delta\kappa_2, - n_3 \, \Delta\kappa_3)^T$. $\Phi_{l,n_1...n_d}^{\alpha}(\theta)$ are $m \times 2^{d-1}$ independent sequences of independent random phase angles drawn at any wave number $\bkappa^{\alpha}_{n_1...n_d}$ from a uniform distribution in the range $[0, \, 2\pi]$.

Random fields generated with relation~\eqref{eq:gVFcomp} are periodic along the $x_i$ axes, $i \in [1,...,d]$, with period:
\begin{equation}\label{eq:Lk}
 L^0_i := \frac{2\pi}{\Delta \kappa_i}
\end{equation}
Also, the values of the field are bounded according to:
\begin{equation}
 g_j(\bx;\theta) \leq 2 \sqrt{\Delta \kappa_1 \cdots \Delta\kappa_d} \ \sum_{l=1}^m \sum_{n_1=0}^{N_1-1} \cdots \sum_{n_d=0}^{N_d-1} \sum_{\alpha=1}^{2^{d-1}} \ \vert H_{jl}(\bkappa^{\alpha}_{n_1...n_d}) \vert 
\end{equation}

It has been shown that the random fields generated according to equation~\eqref{eq:gVFcomp} have the following properties~\cite{ShiDeo1991, ShiDeo1996, Deodatis1996a, Deodatis1996b, PopDeoPre1998}:
\begin{enumerate}
 \item They tend to be standard Gaussian as $N_i \rightarrow +\infty$, $\forall i \in [1,...,d]$; rate of convergence is investigated in~\cite{ShiDeo1991}.
 \item They ensemble auto- and cross-correlations are identical to the target functions.
 \item Each realization is ergodic in mean and correlation (spatial mean and correlation over domain $\mR$ are equal to ensemble mean and correlation) when the size of the spatial domain $\vert \mR \vert$ tends to be infinite in every directions.
 \item Each realization is ergodic in mean as $\vert \mR \vert = L^0_1 \times ... \times L^0_d$ (see equation~\eqref{eq:Lk}).
\end{enumerate}
 For properties 3 and 4 to be true, this further condition has to be satisified: \newline
 $H_{jk}(\kappa_1,..., \kappa_d) = 0$, $(j,k) \in [1,...,m]^2$, as any of the $\kappa_i$, $i \in [1,...,d]$, is equal to zero.
\begin{enumerate}
 \item[5.] As discussed in appendix~1, the random fiels generated from equation~\eqref{eq:gVFcomp} are not ergodic in correlation as $\vert \mR \vert = L^0_1 \times ... \times L^0_d$. However, using properly defined wave-number shifts~\cite{Deodatis1996b, PopDeoPre1998}, ergodicity in correlation is recovered on a finite domain as the spatial correlations are calculated over a domain of size $\vert \mR \vert = m \, L^0_1 \times ... \times m \, L^0_d$. In this case, the wave number vector introduced in equation~\eqref{eq:Dk} is modified so as it also depends on the index $l$, as follows:
 \begin{equation}\label{eq:Delk-shift}
 \bkappa^{\alpha}_{l, n_1...n_d} := \left( (n_1 + \frac{l}{m}) \Delta\kappa_1 \quad I_2^{\alpha} (n_2 + \frac{l}{m}) \Delta\kappa_2 \quad \ldots \quad I_d^{\alpha} (n_d + \frac{l}{m}) \Delta\kappa_d \right)^T
 \end{equation}
\end{enumerate}
Besides, as wave-number shifts are introduced, the condition that functions $H_{jk}(\kappa_1,..., \kappa_d)$ be equal to zero as any $\kappa_i=0$ can be removed for properties 3 and 4 to be valid.

\subsubsection{Translation to non-Gaussian stochastic vector fields}
\label{sec:tra-RF}

The approach presented above generates $m$ zero-mean unit-variance homogeneous Gaussian stochastic fields $\mathfrak{g}_j(\bx; \theta)$, $j \in [1,...m]$, with cross-correlation matrix $\bR^{\bm{g}}(\bxi)$. $m$ homogeneous non-Gaussian stochastic translation fields $\mathfrak{f}_j(\bx; \theta)$ can be obtained from their Gaussian counterparts $\mathfrak{g}_j(\bx; \theta)$. The translation fields are defined by the following memoryless -- meaning that the outputs at any point $\bx$ do not depend on the inputs at any other point -- mapping:
\begin{equation}
 \mathfrak{f}_j(\bx) = \mF_{\mathfrak{f}_j}^{-1}\left( \mF_{\mathfrak{g}_j}(\mathfrak{g}_j(\bx)) \right) = F_j\left(\mathfrak{g}_j(\bx)\right) \quad , \quad j \in [1,...,m]
\end{equation}
where $\mF_{\mathfrak{g}_j}$ is the standard Gaussian cumulative density function (CDF) of the random variables $\mathfrak{g}_j(\bx)$, $\mF_{\mathfrak{f}_j}^{-1}$ the inverse of the marginal CDF of the non-Gaussian random variables $\mathfrak{f}_j(\bx)$, and $F_j = \mF_{\mathfrak{f}_j}^{-1} \circ \mF_{\mathfrak{g}_j}^{}$.

Then, the components of the non-Gaussian correlation matrix can be computed as:
\begin{align} \label{eq:Rf-Rg}
 R^{\bm{f}}_{jk}(\bxi,\brho^{\bm{g}}) :=& \mathbb{E}[\mathfrak{f}_j(\bx) \, \mathfrak{f}_k(\bx+\bxi)]  \nonumber \\
 :=& \int_{-\infty}^{+\infty} \int_{-\infty}^{+\infty} F_j\left(\mathfrak{g}_j(\bx)\right) \, F_k\left(\mathfrak{g}_k(\bx+\bxi)\right) \times \\
 & \qquad\qquad p^G_{\mathfrak{g}_j \mathfrak{g}_k}\left( g_j(\bx), g_k(\bx+\bxi) ; \rho^{\bm{g}}_{jk}(\bxi) \right) \, dg_j(\bx) \, dg_k(\bx+\bxi) \nonumber
\end{align}
where $p^G_{\mathfrak{g}_j \mathfrak{g}_k}$ denotes the standard Gaussian joint PDF of the two random variables $\mathfrak{g}_j(\bx)$ and $\mathfrak{g}_k(\bx+\bxi)$. Note that in the case of standard Gaussian distribution, we have $\rho^{\bm{g}}_{jk}(\bxi) = R^{\bm{g}}_{jk}(\bxi)$ (see equations~\eqref{eq:Rdef0} and~\eqref{eq:def-cor-coeff}).

In practice, one is interested in generating realizations of non-Gaussian random fields $\mathfrak{f}_j$ with targeted marginal PDF $\mF_{\mathfrak{f}_j}^{0}$ and targeted cross-correlation matrix $\bR^{\bm{f} \, 0}(\bxi)$. To that purpose, the cross-correlation matrix $\bR^{\bm{g}}(\bxi)$ of the underlying standard Gaussian fields $\mathfrak{g}_j(\bx; \theta)$ has to be determined. We recall (see equations~\eqref{eq:Rdef0} and~\eqref{eq:def-cor-coeff}) that:
\begin{equation}\label{eq:rho-R}
 \rho^{\bm{f}}_{jk}(\bxi, \rho^{\bm{g}}_{jk}) := \frac{R^{\bm{f}}_{jk}(\bxi, \rho^{\bm{g}}_{jk}) - \mu_{\mathfrak{f}_j} \mu_{\mathfrak{f}_k}}{s_{\mathfrak{f}_j} s_{\mathfrak{f}_k}}
\end{equation}
Suppose that, $\forall (j,k) \in [1,...,m]^2$, we calculate from relations~\eqref{eq:Rf-Rg} and~\eqref{eq:rho-R} the two quantities $\rho_{jk}^{\bm{f} \, min}(\bxi) = \rho^{\bm{f}}_{jk}(\bxi, -1)$ and $\rho_{jk}^{\bm{f} \, max}(\bxi) = \rho^{\bm{f}}_{jk}(\bxi, +1)$. Following~\cite{Grigoriu1995, GioGusGri2000}, if the functions $\rho^{\bm{f}}_{jk}(\bxi)$ all fall in the range $[\rho_{jk}^{\bm{f} \, min}(\bxi), \ \rho_{jk}^{\bm{f} \, max}(\bxi)]$, $\forall \bxi$, then equation~\eqref{eq:Rf-Rg} can be analytically or numerically inverted to calculate a unique $\brho^{\bm{g}}(\bxi)$. Besides, it must be verified that the matrix $\brho^{\bm{g}}(\bxi)$ really is a correlation matrix, namely that the auto-correlation functions $\rho^{\bm{g}}_{jj}(\bxi)$, $j \in [1,...m]$, as well as the correlation matrix $\brho^{\bm{g}}(\bxi)$ are positive semi-definite for every separation distance $\bxi$.

Inverting relation~\eqref{eq:Rf-Rg} is not always possible, and when it is not, cross-correlation matrix $\bR^{\bm{f}}(\bxi)$ and marginal CDFs $\mF_{\mathfrak{f}_j(\bx)}$, $j \in [1,...,m]$, are said to be ``incompatible''~\cite{ShiDeo2013}. In this case, the iterative method presented in~\cite{ShiDeo2013} can be implemented (see also~\cite{BocDeo2008}). With this method, the non-Gaussian CDFs are taken as $\mF_{\mathfrak{f}_j} = \mF^0_{\mathfrak{f}_j}$ and the correlation functions $R^{\bm{g}}_{jk}(\bxi)$ of the underlying standard Gaussian fields are iteratively modified until the correlation fonctions of the translated fields are sufficiently close to the targets: $\bR^{\bm{f}}(\bxi) \approx \bR^{\bm{f} \, 0}(\bxi)$.

\section{Numerical implementation}

\subsection{Random vector fields generation using FFT} 
\label{sec:NumRFgen}

For numerical implementation, equation~\eqref{eq:gVFcomp} is rewritten as:
\begin{align}
 & B^{\alpha}_{jl}(\bkappa^{\alpha}_{n_1...n_d};\theta) := 2 \sqrt{\Delta \kappa_1 \cdots \Delta\kappa_d} \ \vert H_{jl}(\bkappa^{\alpha}_{n_1...n_d}) \vert \, e^{-i \, \varphi_{jl}(\bkappa^{\alpha}_{n_1...n_d})} \, e^{i \, \Phi_{l,n_1...n_d}^{\alpha}(\theta)} \label{eq:Bkl} \\
 & G^{\alpha}_{jl}(\bx_{m_1...m_d} ; \theta) := \sum_{n_1=0}^{N_1-1} \cdots \sum_{n_d=0}^{N_d-1} B^{\alpha}_{jl}(\bkappa^{\alpha}_{n_1...n_d};\theta) \, e^{i \, \bkappa^{\alpha}_{n_1...n_d} \cdot \bx_{m_1...m_d}} \label{eq:gkl} \\
 & g_j(\bx_{m_1...m_d};\theta) = Re \sum_{l=1}^m \sum_{\alpha=1}^{2^{d-1}} G^{\alpha}_{jl}(\bx_{m_1...m_d} ; \theta)  \label{eq:gVFnum}
 \end{align}
for $j \in [1,...,m]$, with $N_i \in \mathbb{N}^{\star}$ and $N_i \rightarrow +\infty$ for all $i \in [1,...,d]$, where $Re(z)$ is the real part of the complex number $z$, and where we introduced:
\begin{equation}\label{eq:Dx}
 \bx_{m_1...m_d} := (m_1 \Delta x_1 \quad m_2 \Delta x_2 \quad \ldots \quad m_d \Delta x_d)^T \quad , \quad m_i \in [0,...,M_i-1]
\end{equation}
Relation~\eqref{eq:gkl} can be numerically computed in an efficient way using fast Fourier transform (FFT) algorithm.

The random fields are generated over a spatial period $L^0_1 \times ... \times L^0_d$ setting
\begin{equation}\label{eq:Lx}
  \Delta x_i := \frac{2\pi}{M_i \, \Delta\kappa_i}
\end{equation}
with $M_i \geq 2 \, N_i$ to avoid aliasing. Introducing definitions~\eqref{eq:Dk} and~\eqref{eq:Dx}, along with~\eqref{eq:Lk} and~\eqref{eq:Lx} in~\eqref{eq:gkl}, and reminding that $B^{\alpha}_{jl}(\bkappa^{\alpha}_{n_1...n_d};\theta)=0$ for any $n_i \geq N_i$, that is $n_i \Delta\kappa_i \geq \kappa_{ui}$, it comes:
\begin{equation}
 G^{\alpha}_{jl}(\bx_{m_1...m_d} ; \theta) := \sum_{n_1=0}^{M_1-1} \cdots \left( \sum_{n_d=0}^{M_d-1} B^{\alpha}_{jl}(\bkappa^{\alpha}_{n_1...n_d};\theta) \, e^{2i\pi \, I^{\alpha}_d \frac{m_d n_d}{M_d}} \right) \cdots e^{ 2i\pi \, I^{\alpha}_1 \frac{m_1 n_1}{M_1}}
\end{equation}
where a sequence of $d$ Fourier or inverse Fourier transforms, according to the sign of $I_i^{\alpha}$, can be recognized.

Note that in the case wave-number shifts are applied, the equations in this section~\ref{sec:NumRFgen} can be straightforwardly adapted by introducing equation~\eqref{eq:Delk-shift} instead of~\eqref{eq:Delk} for the wave numbers vector. The side effect is that the periods over which random fields are generated are elongated as:
\begin{equation}
 L_i^0 = \frac{2\pi}{\Delta \kappa_i} \quad \rightarrow \quad L_i^0 = m \times \frac{2\pi}{\Delta \kappa_i} \quad , \quad i \in [1,...,d]
\end{equation}
and with the random fields generated over the grid (compare to~\eqref{eq:Dx}):
\begin{equation}
 \bx_{m_1...m_d} := (m_1 \Delta x_1 \quad m_2 \Delta x_2 \quad \ldots \quad m_d \Delta x_d)^T \quad , \quad m_i \in [0,...,m \times M_i - 1]
\end{equation}

\subsection{Material response at mesoscale}
\label{sec:MesoMatRes}

The components of the elasticity tensor $\bC$, damage-plasticity ratio $r$ and yield stress $\sigma_y$ are parameters of the material model introduced in section~\ref{sec:MatModMes}. These parameters are realizations of random variables over the ED $\mR$, according to the random vector fields generated as presented in the previous section. For the sake of readability, reference to the spatial position ($\bx$) and to the random experiment ($\theta$) are dropped in this section.

\subsubsection{Discrete evolution equations}

We introduce the discrete process for the pseudo-time: $\mT_0^T = \{ t_n, n \in [0,...,N_T] \}$ with $t_0 = 0$, $t_{N_T} = T$, and the pseudo-time increment $t_{n+1} - t_n := \Delta t$.

The numerical integration of the evolution of the internal variables over the process $\mT_N$ is performed using the unconditionally stable backward Euler time integration scheme. Accordingly, the evolution of the internal variables (see equations~\eqref{eq:D-evo} and~\eqref{eq:epP-evo}) are implemented as:
\begin{align}
 & \bD_{n+1} : \bsigma_{n+1} = \bD_n : \bsigma_{n+1} +r \, \gamma_{n+1} \bnu_{n+1} \label{eq:D} \\
 & \bepsilon^p_{n+1} = \bepsilon^p_n + (1-r) \gamma_{n+1} \bnu_{n+1} \label{eq:eP}
\end{align}
where $\gamma_{n+1} := \gamma(t_{n+1}) := \Delta t \, \dot{\gamma}_{n+1}$.

Besides, considering equation~\eqref{eq:sta-equ}, we have for the stress tensor:
\begin{align}
 & \bD_{n+1} : \bsigma_{n+1} := \bepsilon^d_{n+1} := \bepsilon_{n+1} - \bepsilon^p_{n+1} \\
 \Rightarrow \ & \bD_n : \bsigma_{n+1} = \bepsilon_{n+1} - \bepsilon^p_n - \gamma_{n+1} \bnu_{n+1} \label{eq:sig}
\end{align}

Finally, the tangent modulus in equation~\eqref{eq:tan-mod} is computed as:
\begin{equation}
 \blambda_{n+1} = \left\{
 \begin{array}{ll}
  \bD_{n+1}^{-1} \quad & \textrm{if} \quad \gamma_{n+1} = 0 \quad (\phi(\bsigma_{n+1}) < 0) \\
  \mathbf{0} \quad &  \textrm{if} \quad \gamma_{n+1} > 0 \quad (\phi(\bsigma_{n+1}) = 0 \ ; \ \dot{\phi}(\bsigma_{n+1}) = 0)
 \end{array} \right.
\end{equation}

\subsubsection{Solution procedure}

The problem to be solved at any material point $\bx$ of the mesoscale, reads: \\
\textit{Given $\bepsilon_{n+1} = \bepsilon_n + \Delta\bepsilon_{n+1}$, find $\gamma_{n+1} \bnu_{n+1}$ such that $\phi_{n+1} \leq 0$}. 
This is solved using a so-called return-mapping algorithm (see e.g.~\cite{SimHug1998, Ibrahimbegovic2009}) where a trial state is first considered and followed by a corrective step if required:
\begin{enumerate}
 \item \textbf{Trial state:} \\
 It is assumed that there is no inelastic evolution due to deformation increment $\Delta\bepsilon_{n+1}$, that is $\gamma_{n+1} = 0$. Accordingly, the internal variables remain unchanged: $\bD_{n+1}^{trial} = \bD_n$ and $\bepsilon_{n+1}^{p,trial} = \bepsilon^p_{n}$. The trial stress along with the trial criterium function can then be computed as:
\begin{align}
 & \bsigma^{trial}_{n+1} = \bD^{-1}_n : (\bepsilon_{n+1}-\bepsilon^p_n) \label{eq:sigtr} \\
 & \phi^{trial}_{n+1} = h(\bsigma^{trial}_{n+1}) - \sigma_y \label{eq:phitr}
\end{align}
The admissibility of this trail state then has to be checked:
\begin{itemize}
 \item If $\phi^{trial}_{n+1} \leq 0$, the trial state is admissible and the local variables are updated accordingly: $\bsigma_{n+1} = \bsigma_{n+1}^{trial}$, $\bD_{n+1} = \bD_n$, $\bepsilon^p_{n+1} = \bepsilon^p_n$. Besides, the tangent modulus is: $\blambda_{n+1} = \bD_n^{-1}$. 
 \item If $\phi^{trial}_{n+1} > 0$, the trial state is not admissible and it has to be corrected as described in the next step `correction'.
\end{itemize}

 \item \textbf{Correction:} \\
 If trial state is not admissible, then $\gamma_{n+1} > 0$ and, according to equation~\eqref{eq:loa-unl}, the relation $\phi_{n+1} = 0$ has to be satisfied. Solving $\phi_{n+1} = 0$ yields $\gamma_{n+1} \bnu_{n+1}$.
 
Then, the stresses can be calculated following equation~\eqref{eq:sig}, the internal variables are updated according to equations~\eqref{eq:D} and \eqref{eq:eP}, and finally tangent modulus reads $\blambda_{n+1} = \mathbf{0}$.
 
\end{enumerate}

\subsection{Material response at macroscale}

In this section, we derive the equations to be implemented for the numerical computation of the response of the ED at macroscale, that is the macroscopic behavior law $\dot{\bSigma} = \bL : \dot{\bE}$, where $\bSigma$ and $\bE$ are the macroscopic stress and strain tensors while $\bL$ denotes the homogenized tangent modulus at macroscale.

\subsubsection{Discrete governing equations in the ED}

The weak form of the boundary value problem in equations~\eqref{eq:BVPs} reads:
\begin{align}
 0 &= \int_{\mR} \delta\bu \cdot \mathbf{div} \ \bsigma \ d\mR \nonumber \\
    &= \int_{\mR} \bnabla^s\delta\bu : \bsigma(\bnabla^s\bu) \ d\mR - \int_{\partial\mR} \delta\bu \cdot \bt \ d\partial\mR \label{eq:BVPw}
\end{align}
where $\delta\bu$ is any virtual displacement field that satisfies $\delta\bu = \mathbf{0}$ on $\partial_u\mR$.

Finite element (FE) method is used to approximate the displacement field over the ED. Accordingly, $\mR$ is meshed into $N_{el}$ elements $\mR^e$ such that $\bigcup_{e=1}^{N_{el}} \mR^e = \mR$. Then, in each element, displacement fields is computed as (see e.g.~\cite{ZieTay2000}):
\begin{equation}
 \bu(\bx,t)\vert_{\mR^e} = \bN^e(\bx) \, \bd^e(t)
\end{equation}
where $\bN^e(\bx)$ contains the element shape functions and $\bd^e(t)$ are the displacements at the nodes of the FE mesh. Equation~\eqref{eq:BVPw} can then be rewritten as:
\begin{align} \label{eq:ass-wea-for}
 0 := \overset{N_{el}}{\underset{e=1}{\mathlarger\asmby}} & \left\{ \int_{\mR^e} \mathrm{sym}\left[ \bnabla\left(\bN^e \, \delta\bd^e\right) \right] : \bsigma \left( \mathrm{sym}\left[ \bnabla\left(\bN^e \, \bd^e\right) \right] \right) d\mR^e \right. \nonumber \\ 
 & \qquad\qquad\qquad\qquad\qquad\qquad\qquad  \left. - \int_{\partial\mR^e} \bN^e \, \delta\bd^e \cdot \bt \ d\partial\mR^e \right\}
\end{align}
where $\overset{N_{el}}{\underset{e=1}{\mathlarger\asmby}}$ denotes the finite element assembly procedure, and $\partial\mR^e$ denotes the portion, if any, of the boundary of the element $e$ that is also a part of the boundary of the discretized domain $\partial\mR$.

Matrix notations can be conveniently adopted at this stage, so that the elements of the symmetric second-order tensors $\bsigma$ and $\bepsilon$ are written as vectors:
\begin{equation}
 \bsigma \ \rightarrow \overline{\bsigma} \qquad ; \qquad \bepsilon := \mathrm{sym}\left[ \bnabla\left(\bN^e \, \bd^e\right) \right] \ \rightarrow \ \overline{\bepsilon} := \bB^e \, \bd^e
\end{equation}
The matrix $\bB^e$ is composed of derivatives of the element shape functions. With these notations, we have $\bepsilon : \bsigma \ \rightarrow \overline{\bepsilon}^T \, \overline{\bsigma}$, so that~\eqref{eq:ass-wea-for} can be rewritten as: 
\begin{equation}
 0 := \overset{N_{el}}{\underset{e=1}{\mathlarger\asmby}} \ \delta\bd^{eT} \left\{ \int_{\mR^e} \bB^{eT} \overline{\bsigma}(\overline{\bepsilon}) d\mR^e - \int_{\partial\mR^e} \bN^{eT} \, \bt \ d\partial\mR^e \right\}
\end{equation}

Because the equation above has to be satisfied for any virtual nodal displacements vector $\delta\bd^e$ that satisfies $\delta\bd^e = \mathbf{0}$ at any node on $\partial_u\mR$, this is finally the following set of nonlinear equations that has to be solved:
\begin{equation}
 \mathbf{0} := \br(\bd) := \bF^{int}(\bd) - \bF^{ext}
\end{equation}
where:
\begin{equation}\label{eq:StrucFint}
 \bF^{int}(\bd_{n+1}) := \overset{N_{el}}{\underset{e=1}{\mathlarger\asmby}} \int_{\mR^e} \bB^{eT} \, \overline{\bsigma}_{n+1} \ d\mR^e \quad ; \quad \bF^{ext}_{n+1} := \overset{N_{el}}{\underset{e=1}{\mathlarger\asmby}} \int_{\partial\mR^e} \bN^{eT} \, \bt^e_{n+1} \ d\partial\mR^e
\end{equation}
Here, we added explicit reference to the time discretization to recall that it is a nonlinear evolution problem that has to be solved.

\subsubsection{Solution procedure}
\label{sec:FESolProc}

First, we separate the degrees of freedom of the $N_{bo}$ nodes that are on the boundary $\partial\mR$ of the ED -- denoted by the subscript $\bar{u}$ -- from those pertaining to its interior -- denoted by the subscript $u$ -- and rearrange them so that:
\begin{equation} \label{eq:res0}
 \bd =
 \begin{pmatrix}
  \bd_u \\
  \bd_{\bar{u}}
 \end{pmatrix}
 \qquad \mathrm{and} \qquad
 \br(\bd) =
 \begin{pmatrix}
  \br_u(\bd) \\
  \br_{\bar{u}}(\bd)
 \end{pmatrix} :=
 \begin{pmatrix}
  \bF^{int}_u(\bd) \\
  \bF^{int}_{\bar{u}}(\bd)
 \end{pmatrix} -
 \begin{pmatrix}
  \mathbf{0} \\
  \bF_{\bar{u}}^{ext}
 \end{pmatrix} 
\end{equation}
$\bF_u^{ext} = \mathbf{0}$ because there is no external forces applied on the interior nodes.

As external forces $\bF_{\bar{u}}^{ext}$ increase by $\Delta \bF_{\bar{u}}^{ext}$ and displacements $\bd$ increase by $\Delta\bd$, the residual is linearized such that the problem to be solved now reads:
\begin{equation} \label{eq:lin-set}
 \begin{pmatrix}
  \mathbf{0} \\
  \mathbf{0}
 \end{pmatrix} := 
 \begin{pmatrix}
   \bF_u^{int}(\bd) \\
   \bF_{\bar{u}}^{int}(\bd)
  \end{pmatrix} + 
  \begin{pmatrix}
  \bK_{uu}^{tan} \ & \ \bK_{u\bar{u}}^{tan} \\
  \bK_{\bar{u}u}^{tan} \ & \ \bK_{\bar{u}\bar{u}}^{tan}
 \end{pmatrix}
 \begin{pmatrix}
  \Delta\bd_u \\
  \Delta\bd_{\bar{u}}
 \end{pmatrix} -
 \begin{pmatrix}
  \mathbf{0} \\
  \bF_{\bar{u}}^{ext} + \Delta \bF_{\bar{u}}^{ext}
 \end{pmatrix}
\end{equation}
where $\bK^{tan}$ is the tangent stiffness matrix defined as:
\begin{equation}\label{eq:StrucTanM}
 \bK^{tan} := \frac{\partial \bF^{int}(\bd)}{\partial \bd} = \int_{\mR} \bB^T \, \overline{\blambda} \, \bB \ d\mR = \overset{N_{el}}{\underset{e=1}{\mathlarger\asmby}} \int_{\mR^e} \bB^{eT} \, \overline{\blambda} \, \bB^e \ d\mR^e
\end{equation}
$\overline{\blambda} = \partial\overline{\bsigma} \slash \partial\overline{\bepsilon}$ is the matrix form of the material tangent modulus at mesoscale as introduced in section~\ref{sec:TanMod}. Element tangent stiffness and internal forces vector are numerically computed as:
\begin{align}
 & \bK^{e, tan} := \int_{\mR^e} \bB^{eT} \, \overline{\blambda} \, \bB^e \ d\mR^e \approx \sum_{l=1}^{N_{IP}} \bB^{e T}_l \, \overline{\blambda}_l \, \bB^e_l \, w_l \label{eq:TanStiDis} \\
 & \bF^{e, int} := \int_{\mR^e} \bB^{eT} \, \overline{\bsigma} \ d\mR^e \approx \sum_{l=1}^{N_{IP}} \bB^{eT}_l \, \overline{\bsigma}_l \, w_l \label{eq:IntForDis}
\end{align}
where $w_l$ are the weights associated to the $N_{IP}$ quadrature points.

The macroscopic response of the ED $\mR$ is then computed from its description at mesoscale as follows:
\begin{enumerate}
 \item \textbf{Updating of the imposed displacements on $\partial\mR$:} \\
 Impose displacement $\bd_{\bar{u}} = \bd_{\bar{u}} + \Delta\bd_{\bar{u}}$ on the boundary nodes. We recall that we only consider the case of linear displacements imposed all over the boundary of the ED (see equation~\eqref{eq:hom-BCs}). Following the work presented in~\cite{MieKoc2002, Sav-et-al2014}, we can write these imposed displacements at any node $q$ of the $N_{bo}$ nodes of the boundary as:
\begin{equation}\label{eq:bou-dis}
 \Delta\bd_q = \bW_q^T \, \Delta\overline{\bE} \quad \Rightarrow \quad \Delta\bd_{\bar{u}} = \left[ \bW_1 \ \bW_2 \ \ldots \ \bW_{N_{bo}} \right]^T \Delta\overline{\bE} \ = \ \bW^T  \Delta\overline{\bE}
\end{equation}
where $\overline{\bE}$ is the matrix form of the strain tensor and where the $\bW_q$s are a geometric matrices built from the coordinates $\bx_q$ of the boundary node $q$.

 \item \textbf{Iterative updating of $\Delta\bd_u$:} \\
 Because equations~\eqref{eq:lin-set} are nonlinear, we use Newton-Raphson procedure to iteratively seek $\bd_u$ as $\bd_u^{(k)} = \mathbf{0} + \Delta\bd_u^{(1)} + ... + \Delta\bd_u^{(k)} + ...$ until $\bF_u^{int}(\bd_u^{(l)}) \cdot \Delta\bd_u^{(l)} < tol$ ($\br_u = \bF_u^{int}$). Displacements $\bd_{\bar{u}}$ on the boundary $\partial\mR$ are known from step 1 above, which means that at any iteration $k$, $\Delta\bd_{\bar{u}}^{(k)} = \mathbf{0}$. Then, according to equations~\eqref{eq:lin-set}, we have at every iteration:
\begin{equation}
 \Delta\bd_{u}^{(k+1)} = - \left( \bK^{tan,(k)}_{uu} \right)^{-1} \, \bF^{int}_u(\bd^{(k)})
 \end{equation}

 \item \textbf{Compute stresses at macroscale:} \\
 First, the external forces vectors $\bF^{ext}_q$ (reactions) at the nodes of the boundary are retrieved as:
\begin{equation}
 \br_{\bar{u}}(\bd^{(l)}) := \mathbf{0} \quad \Rightarrow \quad \bF_q^{ext} = \bF^{int}_q(\bd^{(l)}) \quad , \quad q \in [1,...,N_{bo}]
 \end{equation}
Then, the approximation $\bt(\bx_q) d\partial\mR \approx \bF_q^{ext}$ is introduced in equation~\eqref{eq:Sig-from-t} and consider the matrix form $\overline{\bSigma}$ of the stress tensor, which yields~\cite{MieKoc2002}:
\begin{equation}\label{eq:mac-sig}
 \Delta \overline{\bSigma} = \frac{1}{\vert \mR \vert} \sum_{q=1}^{N_{bo}} \bW_q \, \Delta\bF^{ext}_q = \frac{1}{\vert \mR \vert} \bW \, \Delta\bF^{ext}_{\bar{u}}
\end{equation}
  
 \item \textbf{Compute tangent modulus at macroscale:} \\
Considering an equilibrium state, we have $\bF_u^{int}(\bd) = \mathbf{0}$ and $\bF_{\bar{u}}^{int}(\bd) = \bF_{\bar{u}}^{ext}$. Then, according to equations~\eqref{eq:lin-set}, it comes:
\begin{equation} \label{eq:del-Fext}
 \Delta\bd_{u} = - \left( \bK^{tan}_{uu} \right)^{-1} \, \bK^{tan}_{u\bar{u}} \, \Delta\bd_{\bar{u}} \quad
 \Rightarrow \quad \Delta\bF^{ext}_{\bar{u}} = \tilde{\bK}^{tan}_{\bar{u}\bar{u}} \, \Delta\bd_{\bar{u}}
\end{equation}
where $\tilde{\bK}_{\bar{u}\bar{u}} = \bK_{\bar{u}\bar{u}} - \bK_{\bar{u}u} \, \bK^{-1}_{uu} \, \bK_{u\bar{u}}$. Now, combining equations~\eqref{eq:mac-sig}, \eqref{eq:del-Fext} and~\eqref{eq:bou-dis}, it comes the following expression for the matrix form of the tangent modulus at macroscale:
\begin{equation}
 \overline{\bL} := \frac{\Delta \overline{\bSigma}}{\Delta \overline{\bE}} = \frac{1}{\vert \mR \vert} \bW \, \tilde{\bK}_{\bar{u}\bar{u}} \, \bW^T
 \end{equation}
 
\end{enumerate}

It has to be reminded that the macroscopic stresses and tangent moduli are computed for a given realization $\theta$ of the random fields: we have $\overline{\bSigma} = \overline{\bSigma}(\bX, \theta)$ and $\overline{\bL} = \overline{\bL}(\bX, \theta)$. Consequently, there is no guarantee at this point that these quantities are representative of the macroscopic behavior of the material. However, it will be shown in the numerical applications below that for particular structures of the random vector fields that describe an equivalent mesoscale for the material, these macroscopic quantities are almost independent of the realization of the vector fields.

\section{Numerical applications}
\label{sec:NumApp}

The purpose of the following numerical applications is twofold. i) It is first demonstrated in this section that the random vector fields can be parameterized such that a homogeneous material response can be retrieved at macroscale without stochastic homogenization. In this case, macroscopic response does not depend on the realization of the random vector fields that represent variability at an underlying equivalent heterogeneous mesoscale: any realization of the meso-structure yields the same macroscopic response. Consequently, the computational effort is contained at the mesoscale where the nonlinear response of numerous material points has to be computed. ii) We remind that, because only homogeneous displacement boundary conditions are considered in this work, the homogenous response so retrieved at macroscale is a priori dependent on the boundary conditions. This issue is out of the scope here where we focus on showing that the proposed approach can represent salient features of the concrete macroscopic response in compressive cyclic loading while such features are not explicitly present at the mesoscale (emergence of a macroscopic response).

\subsection{1D homogenized response at macroscale}

Throughout this section, we only consider uni-dimensional (1D) material behavior in uniaxial loading at any point $\bX$ of the macroscale. Consequently, strain and stress vectors $\overline{\bE}(\bX,t)$ and $\overline{\bSigma}(\bX,t)$ degenerate into scalar quantities, respectively $\overline{E}_{33} := \overline{E}$ and $\overline{\Sigma}_{33} := \overline{\Sigma}$.

\subsubsection{Spatial discretization at E-mesoscale}

\begin{figure}
  \includegraphics[width=1.0\textwidth]{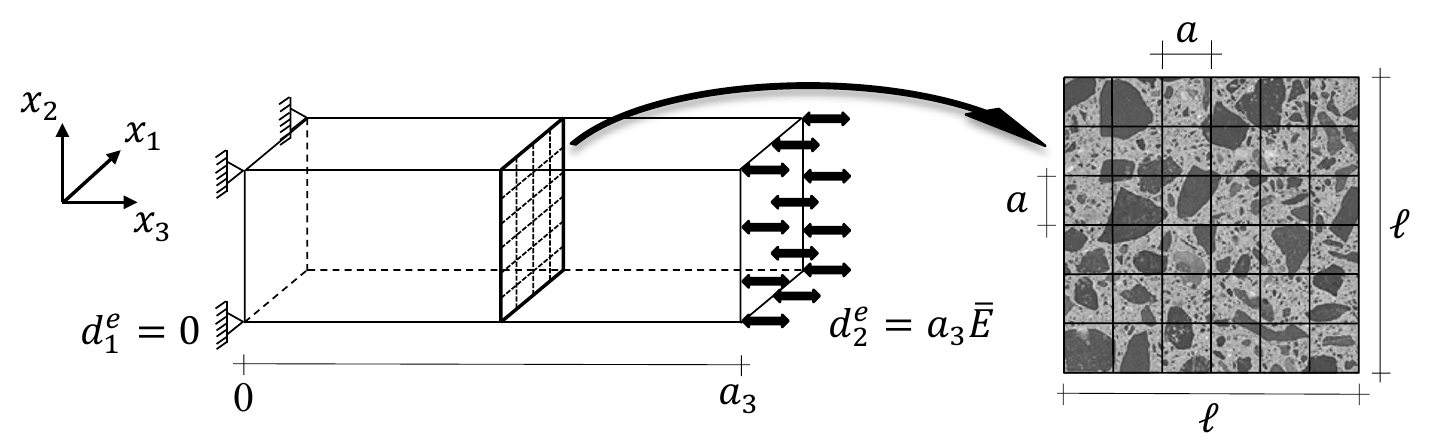}
 \caption{Concrete ED $\mR$ at material point $\bX$ of the macroscale. 1D material response only is considered. ED is discretized into $N_{el} = M_f \times M_f$ adjacent bar elements of length $a_3$ and cross-section $a \times a$. Zero displacement is imposed on the left-hand boundary ($x_3=0$) and homogeneous displacement $\bar{u} = a_3 \overline{E}$ is imposed all over the right-hand boundary ($x_3=a_3$) along the $x_3$-axis.}
 \label{fig:NumApp}
\end{figure}

Accordingly, elementary domain (ED) $\mR(\bX)$ is discretized in the framework of the Finite Element (FE) method as a series of $N_{el}$ adjacent two-node bar elements as shown in figure~\ref{fig:NumApp}. The elements do not have common nodes, they are connected through the boundary conditions at $x_3 = 0$ and $x_3 = a_3$. Each node of the FE mesh has one degree of freedom along $x_3$-axis; besides, each of these nodes belongs to the boundary $\partial\mR$ of the ED, that is $\bd_{\bar{u}} = \bd$ where:
\begin{equation}
 \bd := \left( d_1^1 \ d_2^1 \ \ldots \ d_1^{N_{el}} \ d_2^{N_{el}} \right)^T
\end{equation}
Then, homogeneous kinematic boundary conditions are imposed such that :
\begin{equation}
 \Delta\bd = \bW^T \, \Delta \overline{E} \qquad \mathrm{with} \qquad \bW = \left( 0 \ a_3 \, \ldots \, 0 \ a_3 \right)
\end{equation}

At the bar element level: $\bd^e = (d_1^e \ d_2^e)^T = (0 \ a_3)^T \, \overline{E}$, $\forall e \in [1,...,N_{el}]$. Besides, the shape functions are: 
\begin{equation}
 \bN^e = \left(1 - \frac{x_3}{a_3} \quad \frac{x_3}{a_3}\right) \qquad \Rightarrow \qquad \bB^e = \left(- \frac{1}{a_3} \quad \frac{1}{a_3}\right)
\end{equation}

Also, only one numerical integration point is considered along the $x_3$-axis in each bar element. This implies that the heterogeneity of material properties only has to be accounted for over an ED cross-section and not all over the 3D domain $\mR$. The size of $\mR$ is $\vert \mR \vert = \ell \times \ell \times a_3$ and bar elements $\mR^e$ are assumed to all have the same size:
\begin{equation}
 \vert \mR^e \vert = a \times a \times a_3 \qquad \mathrm{with} \qquad a := \frac{\ell}{M_f} \quad , \quad M_f \in \mathbb{N}^{\star}
\end{equation}

\subsubsection{Homogenized response}

FE approximation introduced above yields, $\forall e \in [1,...,N_{el}]$:
\begin{equation}
\overline{\epsilon}^e := \bB^e \, \bd^e = \overline{E}
\end{equation}
Also, tangent stiffness matrix and internal forces vector in equations~\eqref{eq:StrucTanM} and~\eqref{eq:StrucFint} reads:
\begin{equation}
 \bK^{tan} =
  \begin{pmatrix}
   \bK^{1, tan} &           & \mathbf{0}    \\
                      & \ddots &                     \\
   \mathbf{0}   &            & \bK^{N_{el}, tan}  
  \end{pmatrix}
 \qquad \mathrm{and} \qquad \bF^{e, int} =
 \begin{pmatrix}
   \bF^{1, int} \\
   \vdots \\
   \bF^{N_{el}, int}
 \end{pmatrix}
\end{equation}
with, $\forall e \in [1,...,N_{el}]$:
\begin{equation}
 \bK^{e, tan} = \frac{a^2 \, \overline{\lambda}^e}{a_3}
  \begin{pmatrix}
   1 & -1 \\
   -1 & 1
  \end{pmatrix}
 \qquad \mathrm{and} \qquad \bF^{e, int} = a^2 \, \overline{\sigma}^e
 \begin{pmatrix}
   -1 \\
   1
 \end{pmatrix}
\end{equation}
where $\overline{\lambda}^e$ and $\overline{\sigma}^e$ are the tangent modulus and stress computed at the numerical integration point in any bar element $e$ given $\overline{\epsilon}^e = \overline{E}$ according to the procedure described in section~\ref{sec:MesoMatRes}. Finally, the homogenized quantities at macroscale can be computed as:
\begin{equation}
 \overline{\Sigma} = \frac{1}{M_f^2} \ \sum_{e=1}^{N_{el}} \ \overline{\sigma}^e \qquad \mathrm{and} \qquad \overline{L} = \frac{1}{M_f^2} \ \sum_{e=1}^{N_{el}} \ \overline{\lambda}^e
\end{equation}

\subsection{Heterogeneous structure at E-mesoscale}

In this section, it is described how information is transferred from A-mesoscale to E-mesoscale.

\subsubsection{Assumptions about the structure of the random vector fields}

The following assumptions, previously introduced in~\cite{PopDeoPre1998} for modeling material properties, significantly simplify the equations introduced in section~\ref{sec:NumRFgen}:
\begin{itemize}
 \item The fields have quadrant symmetry, which implies that the cross-correlation matrix is symmetric and real ($\varphi_{jl}(\bkappa^{\alpha}_{n_1n_2}) = 0$, $\forall (j , l)$ and $\forall \bkappa^{\alpha}_{n_1n_2}$);
 \item The auto-correlation functions $R^0_{jk} = R^0$ are identical for every components of the vector field;
 \item The cross-correlation functions $R^0_{jk}$, $j \neq k$ are expressed as $R^0_{jk} = \rho_{jk} R^0$, where the $\rho_{jk}$, $j,k=1,2,3$, are so-called correlation coefficients between the components $C$, $\sigma_y$ and $r$ of the random vector field. They satisfy $-1 \leq \rho_{jk} \leq 1$.
\end{itemize}

Accordingly, cross-spectral density matrix of the random vector field reads:
\begin{equation} \label{eq:CorrCoeff}
 \mathbf{S}^0(\kappa_1, \kappa_2) = S^0(\kappa_1, \kappa_2) \ \bs \quad \textrm{with} \quad \bs =
 \begin{pmatrix}
  1 & \rho_{12} & \rho_{13} \\
  \rho_{12} & 1 & \rho_{23} \\
  \rho_{13} & \rho_{23} & 1
 \end{pmatrix}
\end{equation}
and Cholesky's decomposition can be applied to $\bs$ yielding:
\begin{equation}
 \bs = \bh \, \bh^T
\end{equation}
where $\bh$ is a lower triangular matrix. Then, matrix $\bH$ (see equation~\eqref{eq:Bkl} for instance), reads:
\begin{equation}
 \bH(\kappa_1, \kappa_2) = \sqrt{S^0(\kappa_1, \kappa_2)} \ \bh
\end{equation}

For autocorrelation function, we choose the following form:
\begin{equation} \label{eq:R0-NA}
 R^0(\overline{\xi}_1, \overline{\xi}_2) = s^2 \, \exp \left( - \left( \frac{\overline{\xi}_1}{\overline{b}_1} \right)^2 - \left( \frac{\overline{\xi}_2}{\overline{b}_2} \right)^2 \right)
\end{equation}
where $s^2$ is the variance of the stochastic fields, $\overline{\xi}=\xi \slash \ell$, $\overline{b} = b \slash \ell$ is proportional to $\overline{\ell_c} = \ell_c \slash \ell$ with $\ell_c$ denoting the so-called correlation length.

Appealing to the Wiener-Khinchin theorem, we have the power spectral density function that corresponds to the Fourier transform of the correlation function:
\begin{equation} \label{eq:S0-NA}
 S^0(\overline{\kappa}_1, \overline{\kappa}_2) = s^2 \frac{\overline{b}_1 \overline{b}_2}{4\pi} \exp \left( - \left( \frac{\overline{b}_1 \overline{\kappa}_1}{2} \right)^2 - \left( \frac{\overline{b}_2 \overline{\kappa}_2}{2} \right)^2 \right)
\end{equation}
where $\overline{\kappa} = \kappa \times \ell$.

We define in a general way the correlation length $\overline{\ell_c}$ as the distance such that $R(\overline{\ell_c}) \leq \epsilon_R \, R(0)$ with $0 < \epsilon_R << 1$. From equation~\eqref{eq:R0-NA} comes:
\begin{equation}\label{eq:Cor-b-lc}
 \overline{\ell_c} = \overline{b} \, \sqrt{\ln \frac{1}{\epsilon_R}}
\end{equation}
which clearly shows how parameter $b$ is related to the correlation length $\ell_c$.

Finally, we analogously define the cut-off wave number as $S(\kappa_u) \leq \epsilon_S \, S(0)$ with $0 < \epsilon_S << 1$, which from equation~\eqref{eq:S0-NA} leads to:
\begin{equation}\label{eq:ku1}
 \vert \overline{\kappa}_u \vert \geq \frac{2}{\overline{b}} \sqrt{\ln \frac{1}{\epsilon_S}}
\end{equation}

\subsubsection{Parameterization for random field discretization}

For the numerical applications, random vector fields are generated according to equation~\eqref{eq:gVFnum} with wave-number shifts introduced as in equation~\eqref{eq:Delk-shift}.

Hereafter, random fields parameterization is the same in both directions: $N_1 = N_2 = N$, $M_1 = M_2 = M$, $\kappa_{u \, 1} = \kappa_{u \, 2} = \kappa_u$, and $\ell_{c \, 1} = \ell_{c \, 2} = \ell_c$. Then, the random fields are periodic along $x_1$- and $x_2$-axes with same period:
\begin{equation} \label{eq:L0bar1}
 L^0 := m \frac{2\pi \, N}{\kappa_u} \qquad \mathrm{or} \qquad \overline{L^0} := m \frac{2\pi \, N}{\overline{\kappa}_u}
\end{equation}
with the dimensionless quantities $\overline{L^0} = L^0 \slash \ell$ and $\overline{\kappa}_u = \kappa \times \ell$. Also, random fields are digitized into $m \cdot M \times m \cdot M$ points regularly distributed over a square grid of size $L^0 \times L^0$. The distance between two adjacent points in both directions of the grid is $\Delta x = L^0 \slash (m \, M)$, or $\overline{\Delta x} = \overline{L^0} \slash (m \, M)$ where $\overline{\Delta x} = \Delta x \slash \ell$.

To define a straightforward mapping of the random field grid onto the FE mesh over $\mathcal{R}$, we set:
\begin{equation} \label{eq:L0bar2}
 \Delta x = a \qquad \Rightarrow \qquad \overline{\Delta x} = \frac{1}{M_f} \qquad \Rightarrow \qquad  \overline{L^0} = \frac{m \, M}{M_f}
\end{equation}
Also, we enforce the following condition to avoid any situation where the random material meso-structure would show some periodicity:
\begin{equation}\label{eq:Cond-MMf}
 \overline{L^0} \geq 1 \qquad \Rightarrow \qquad m \, M \geq M_f
\end{equation}
Then, combining equations~\eqref{eq:L0bar1} and~\eqref{eq:L0bar2}, we have:
\begin{equation}\label{eq:ku2}
 \overline{\kappa}_u = 2 \pi \frac{N}{m \, M} M_f
\end{equation}
which introduced in relation~\eqref{eq:ku1} yields:
\begin{equation}\label{eq:Cond-MN}
 \frac{M_f}{m \, M} \geq \frac{1}{\pi \, N \overline{b}} \sqrt{\ln \frac{1}{\epsilon_S}}
\end{equation}
Finally, recalling equation~\eqref{eq:Cor-b-lc}, we have the following relations that the parameterization has to satisfy:
\begin{equation}\label{eq:RelForPar}
 1 \ \geq \ \frac{M_f}{m \, M} \ \geq \ \frac{1}{\pi \, N \overline{\ell_c}} \ln \frac{1}{\epsilon_{RS}} 
\end{equation}
with $0 < \epsilon_{RS} = \epsilon_R = \epsilon_S << 1$.

For all the numerical applications presented hereafter, we choose $M_f = 96$, $N=$~16 and $\epsilon_{RS} = 0.01$. With this parameterization, $\overline{\Delta x} = 1 \slash M_f = 0.010$. Then, to avoid aliasing in the computation of the FFTs (see section~\ref{sec:NumRFgen}), we take $M \geq 2 \, N \geq 32$. With this choice, $m \, M \geq 96 \geq M_f$ so that the left-hand part in~\eqref{eq:RelForPar} is satisfied. The right-hand part in~\eqref{eq:RelForPar} can be rewritten as:
\begin{equation}\label{eq:LcMin}
 \overline{\ell_c} \geq \frac{m \, M}{\pi \, N \, M_f} \ln \frac{1}{\epsilon_{RS}} = \overline{\ell_c}_{min}
\end{equation}
meaning that this parameterization in not appropriate for all correlation lengths.

\subsubsection{Parameterization of the 1D material response at E-mesoscale}

The material law $\Delta\overline{\sigma}^e = \overline{\lambda}^e \, \Delta\overline{\epsilon}^e$ considered in these numerical applications corresponds to the 1D version of the equations developed in section~\ref{sec:MatModMes} completed by the set of equations in appendix~2. Figure~\ref{fig:pnt-law} shows cyclic compressive response obtained from this model at two material points of E-mesoscale, that is in two different elements of the FE mesh over the elementary domain $\mR$.

In each element $e$ of the FE mesh over $\mR$, material parameters $C^e$, $\sigma^e_y$ and $r^e$ take different values due to material heterogeneities. The spatial variability of these three parameters ($m = 3$) over any cross-section of $\mR$ (d=2) is represented by a 3-variate 2-dimensional random vector field that is generated following section~\ref{sec:NumRFgen} with wave-number shifts introduced.

Correlation coefficients in equation~\eqref{eq:CorrCoeff} are set to $\rho_{12} = \rho_{13} = \rho_{23} = 0.9$. This corresponds to strongly correlated random fields, which comes from considering that the three parameters all depend on the geometrical structure of concrete at A-mesoscale: aggregates in a hardened cement paste, as illustrated in figure~\ref{fig:dam-pla-r}.

In the absence of experimental evidence about A-mesoscale, we choose uniform distributions for the parameters, except for the elastic modulus. The reason why a log-normal distribution has been retained for $C$ will be apparent in section~\ref{sec:EmeComRes}. Specifically, table~\ref{tab:DisLaw} presents the distributions used hereafter to build an E-mesoscale that would yield a macroscopic response exhibiting salient features of concrete 1D response in uniaxial compressive cyclic loading. How to translate Gaussian fields to uniform fields is described in appendix~3.

\begin{table}
\caption{Distribution laws for the set of heterogeneous material parameters}
\label{tab:DisLaw}
\begin{tabular}{p{1.5cm}p{3cm}p{2.5cm}p{2.5cm}p{1.5cm}}
\hline\noalign{\smallskip}
 Parameter & Distribution law & Mean & Std. deviation & $COV$ \\
 \hline
 $C$ & $log \mathcal{N}(30e^3, 15e^3)$ & $\mu_C = 30.0$~GPa & $s_C = 15.0$~GPa & 50.0\% \\
 $\sigma_y$ & $\mathcal{U}(0, 70)$ & $\mu_{\sigma_y} = 35.0$~MPa & $s_{\sigma_y} = 20.2$~MPa & 57.7\% \\
 $r$ & $\mathcal{U}(0, 0.6)$ & $\mu_r = 0.3$ & $s_r = 0.17$ & 56.7\% \\
\hline
\end{tabular}
\end{table}


\subsection{Concrete response in uniaxial compressive cyclic loading}

Based on the preceding assumptions and equations, 1D macroscopic response of concrete in uniaxial compressive cyclic loading is now numerically computed. The general-purpose Finite Element Analysis Program FEAP~\cite{FEAP2005} is used for the finite element solution procedure; Python~\cite{Python343} has been used for the implementation of the equations to generate random vector fields; a Python interface has been developed to both generate the random fields and run the FE analyses in an automatic procedure.

The purpose here is not to present a parametric analysis but to show that with the proposed approach, homogeneous response can be retrieved at macroscale without stochastic homogenization and to show that characteristic features of the concrete uniaxial response in cyclic compressive loading at macroscale can emerge from numerous simpler correlated nonlinear and uncertain mechanisms at E-mesoscale. More details about the potential influence of random field properties on the stochastic finite element method, albeit not exactly in the same context as the work presented here, can be found for instance in~\cite{ChaSchPel2007}.

Also, in the absence of detailed information about the correlations at E-mesoscale, the potential problem of incompatible correlation matrix and marginal CDFs presented in section~\ref{sec:tra-RF} has not been treated in these numerical applications.

\subsubsection{Modeling concrete representative surface element}
\label{sec:ModConRep}

We first investigate whether a representative response of the concrete area $\mR$ can be retrieved at macroscale by the proposed modeling. It is reminded that only one type of boundary conditions is considered in this work, namely homogeneous displacements. Consequently, the results shown hereafter could be different for other boundary conditions and the terms ``representative response'' have to be interpreted accordingly.

\begin{figure}
 \begin{tabular}{c}
  \includegraphics[width=1.0\textwidth]{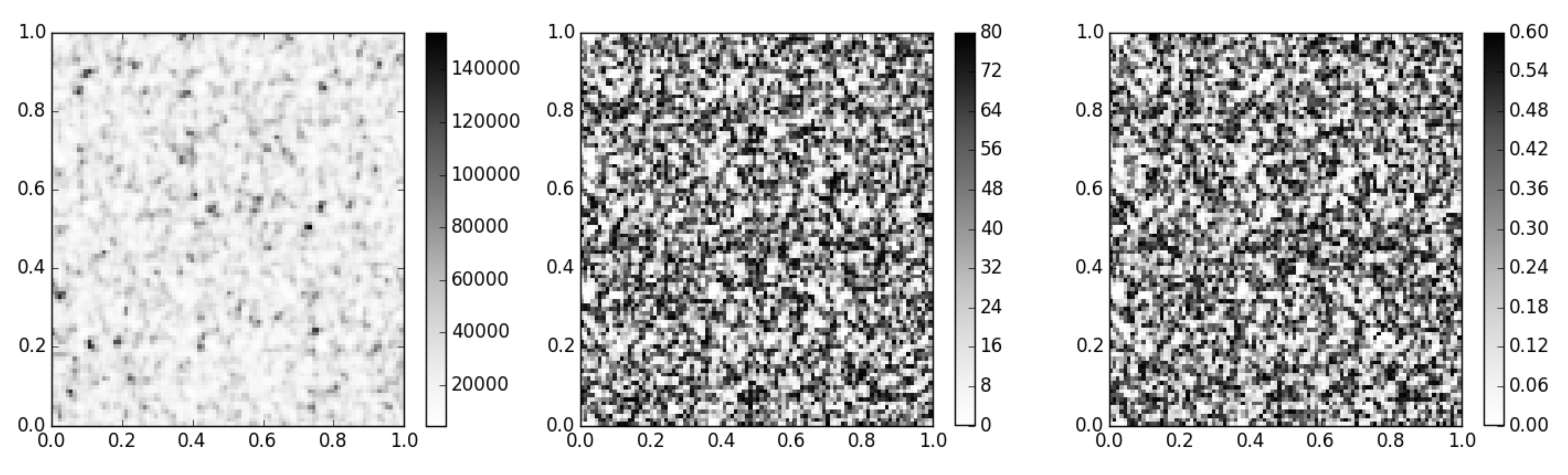} \\
  \includegraphics[width=1.0\textwidth]{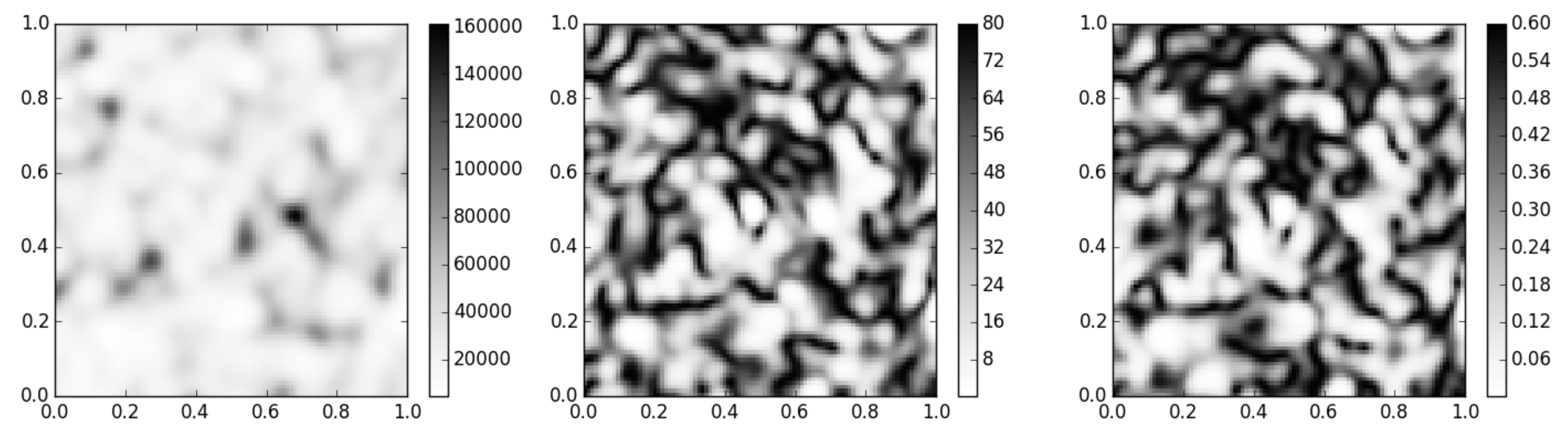}
 \end{tabular}
 \caption{Samples of heterogeneous meso-structures generated over a normalized area $\overline{\mR} = \{ (\overline{x}_1, \, \overline{x}_2) \in [0, \, 1]^2\}$ meshed into $M_f \times M_f = 96 \times 96$ squares and with $M = 32$. [top] $\overline{\ell_c} =$~0.1 (case \#3); [bottom] $\overline{\ell_c} =$~0.4 (case \#1); [left] Elastic modulus $C$ [MPa]; [middle] Yield stress $\sigma_y$  [MPa]; [right] Damage-plasticity coupling ratio $r$ [-].}
 \label{fig:RFs}
\end{figure}

\begin{figure}
 \begin{tabular}{cc}
 \includegraphics[width=0.5\textwidth]{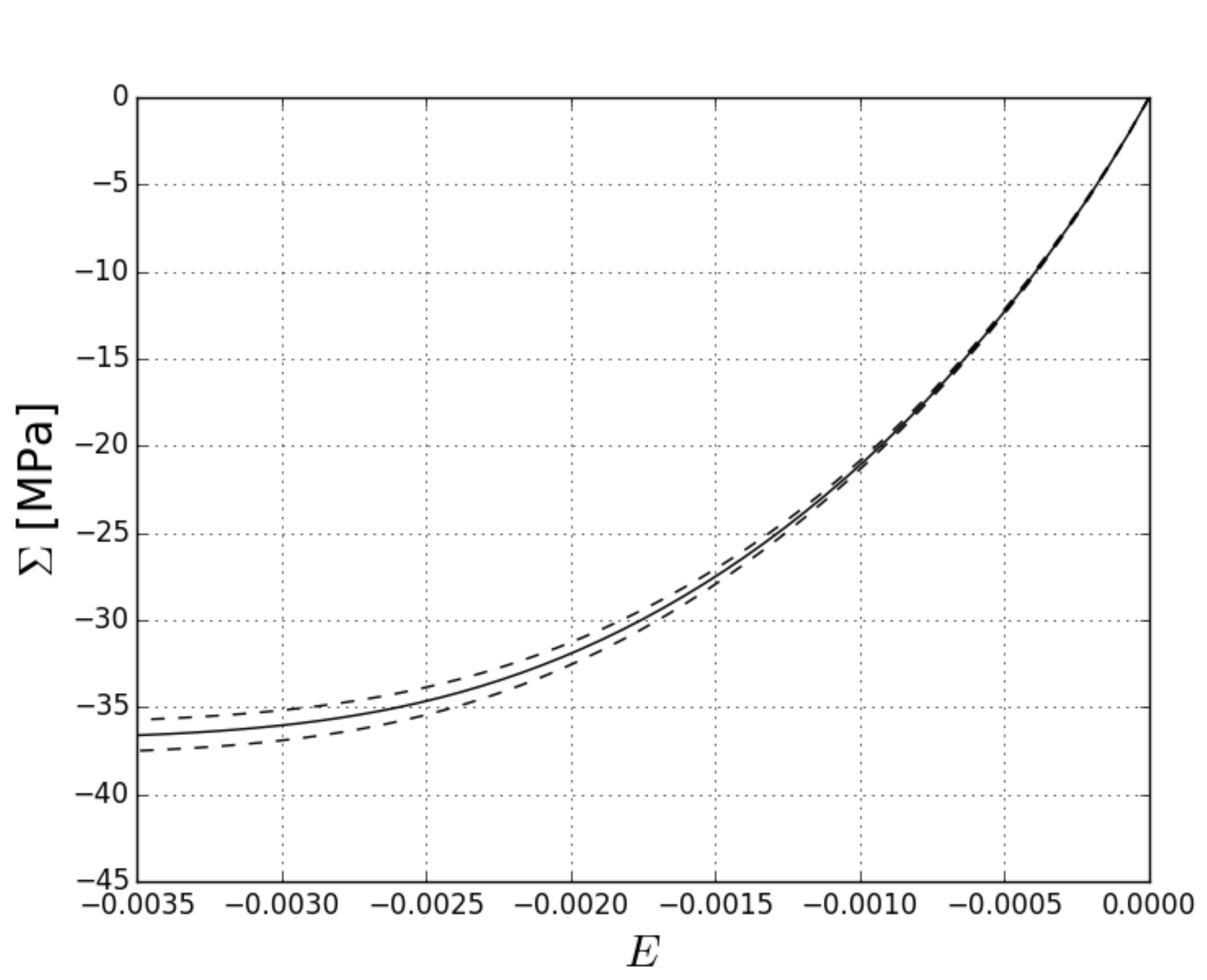} & \includegraphics[width=0.5\textwidth]{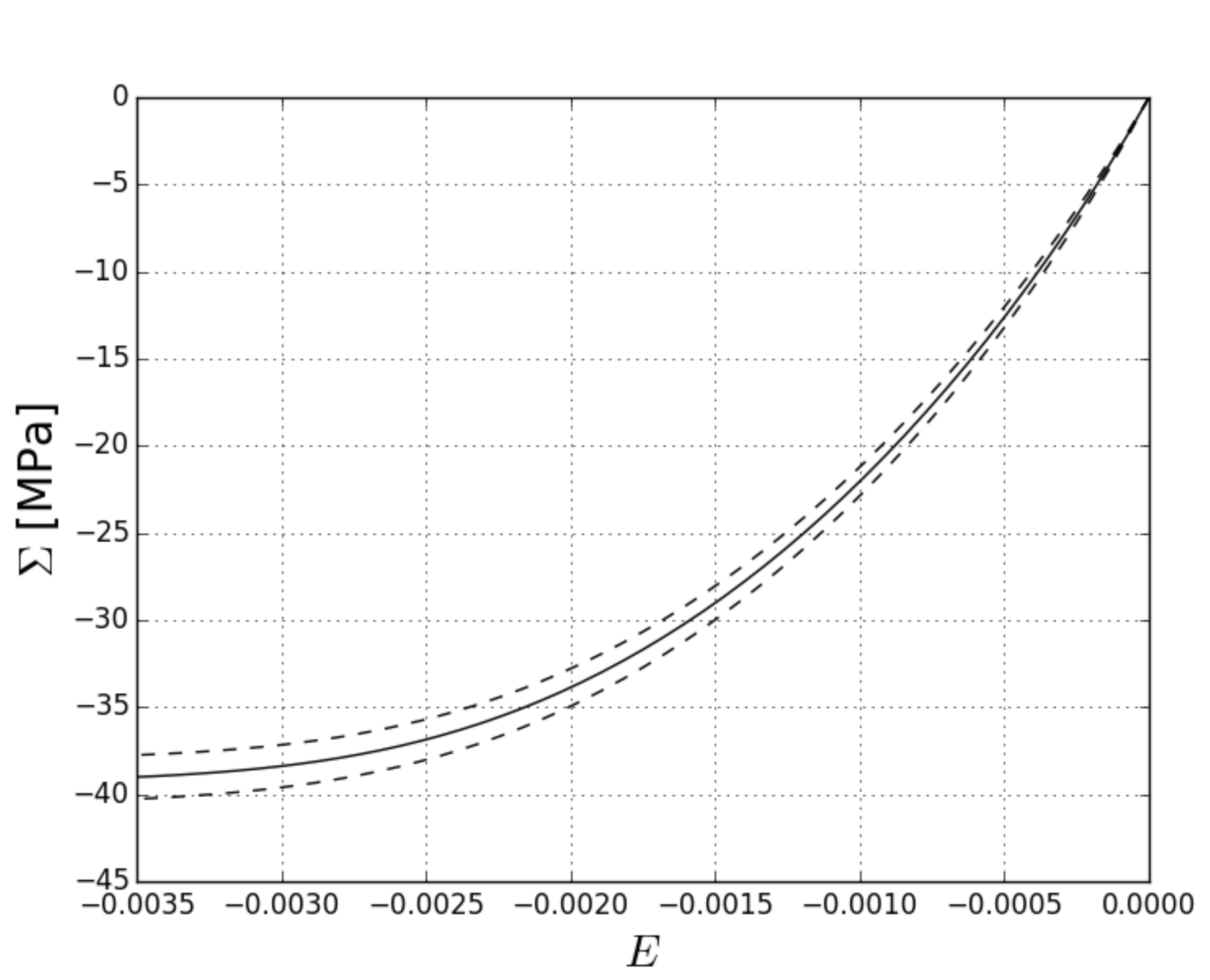} \\
 Case \#1 & Case \#5 \\
 \includegraphics[width=0.5\textwidth]{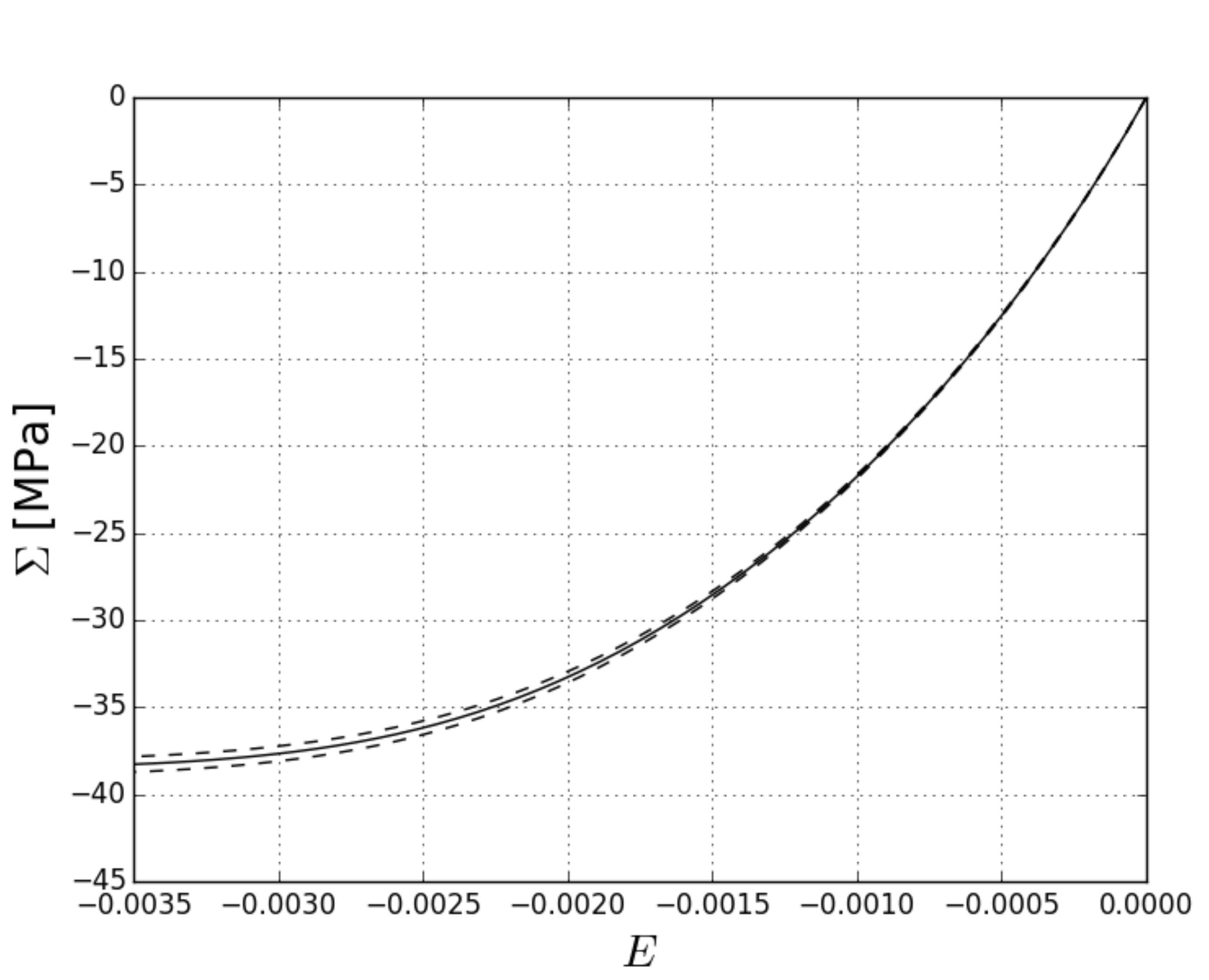} & \includegraphics[width=0.5\textwidth]{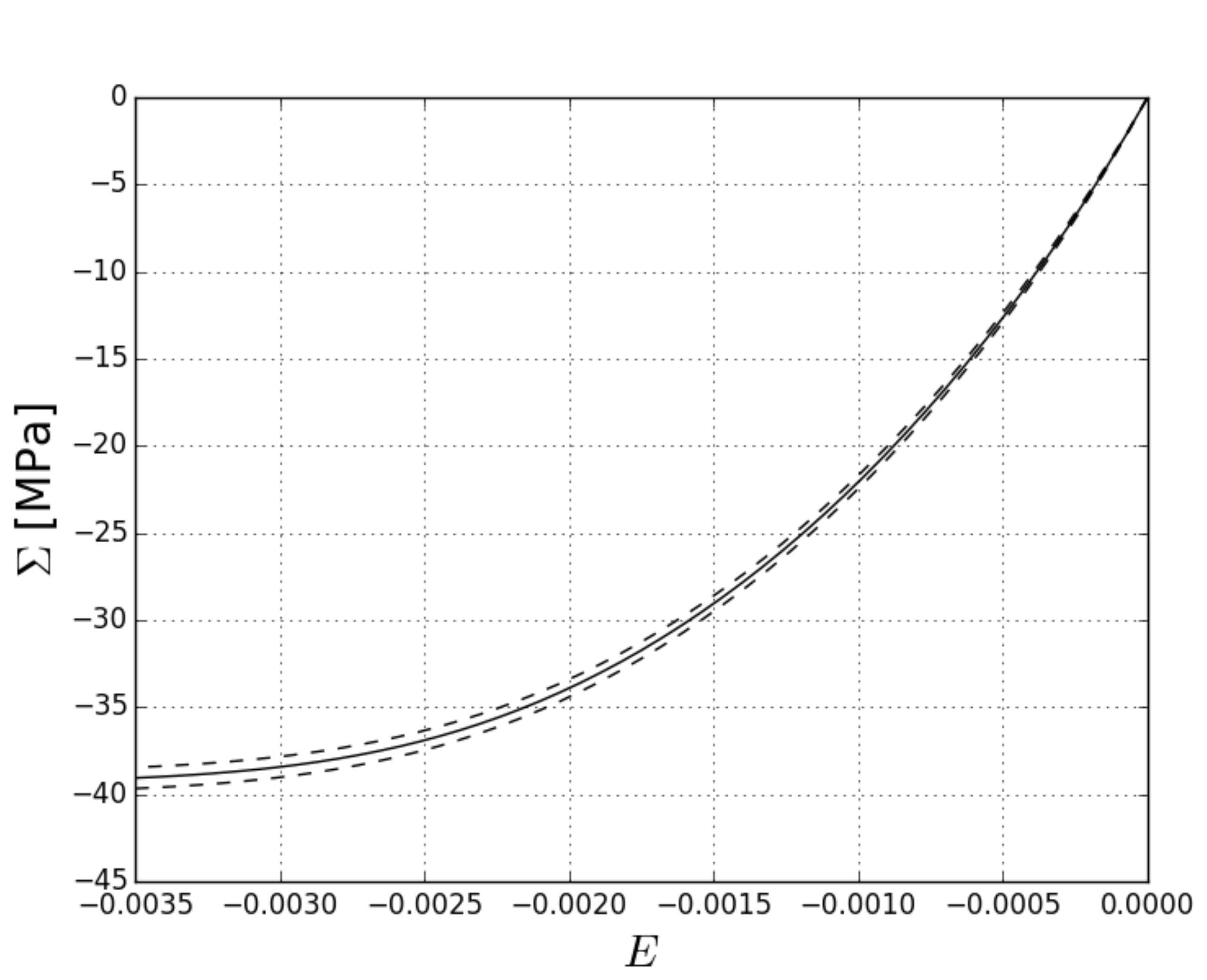} \\
 Case \#2 & Case \#4 \\
 \includegraphics[width=0.5\textwidth]{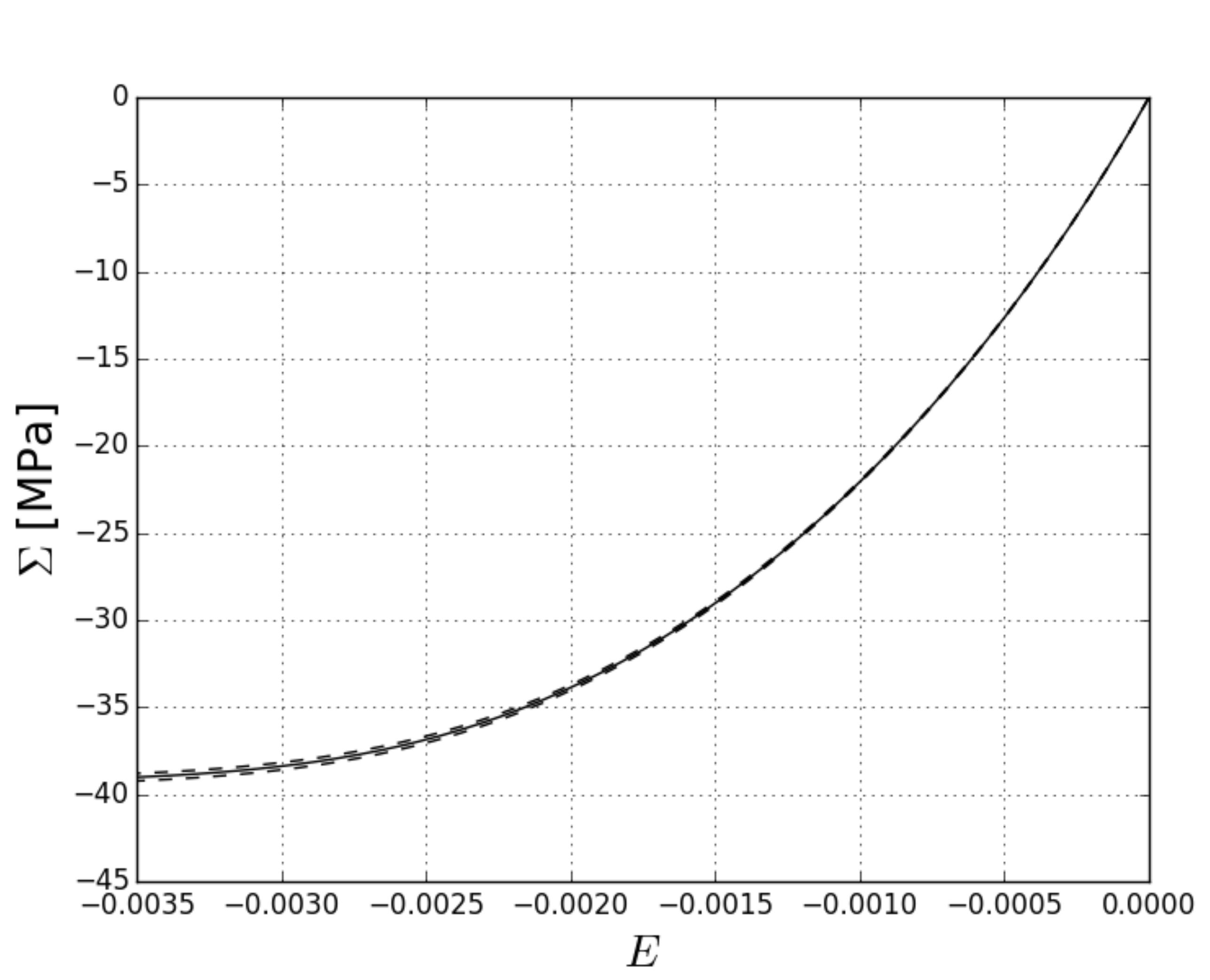} & \includegraphics[width=0.5\textwidth]{Sig-Eps_M32-Lc01.pdf} \\
 Case \#3 & Case \#3
 \end{tabular}
 \caption{Mean (solid line) along with mean plus or minus standard deviation (dashed lines) of the 500 1D macroscopic responses of the ED $\mR$ in uniaxial monotonic compression for the 5 meso-structures considered (cases \#1 to \#5).}
\label{fig:ConLaw-1}
\end{figure}

5 different combinations of parameter $\overline{L^0}$ and correlation length $\overline{\ell_c}$ are considered (see table~\ref{tab:StatResp}). 500 realizations of meso-structures are generated for each of these 5 cases. Figure~\ref{fig:RFs} shows samples of such meso-structures in cases \#1 and \#3. The 500 corresponding 1D macroscopic responses in uniaxial monotonic compressive loading are computed for each of the 5 cases. Figure~\ref{fig:ConLaw-1} presents the mean and standard deviation of these macroscopic responses ($\Sigma$-$E$ law) throughout loading evolution. 

\begin{table}
\caption{500 meso-structures are generated for 5 parameterizations. The mean $\mu$, standard deviation $s$ and coefficient of variation $COV$ of the macroscopic responses of the concrete ED $\mR$ are computed at the end of the monotonic compressive loading ($E = -3.5e^{-3}$).}
\label{tab:StatResp}
\begin{tabular}{p{1.5cm}p{1cm}p{1cm}p{1cm}p{1.5cm}p{1.5cm}p{1.5cm}p{1.5cm}}
\hline\noalign{\smallskip}
Case \# & $M$ & $\overline{L^0}$ & $\overline{\ell_c}$ & $\overline{\ell_c}_{min}$ & $\mu_{3.5}$ [MPa] & $s_{3.5}$ [MPa] & $COV_{3.5}$ [\%]  \\
\hline
1 & 32 & \textbf{1} & 0.4 & 0.09 & -36.6 & 0.89 & 2.4 \\
2 & 32 & \textbf{1} & 0.2 & 0.09 & -38.3 & 0.44 & 1.2 \\
\textbf{3} & \textbf{32} & \textbf{1} & \textbf{0.1} & \textbf{0.09} & \textbf{-39.1} & \textbf{0.22} & \textbf{0.6} \\
4 & 64   & 2 & 0.2 & 0.18 & \textbf{-39.1} & 0.61 & 1.6 \\
5 & 128 & 4 & 0.4 & 0.37 & \textbf{-39.0} & 0.22 & 3.3 \\
\hline
\end{tabular}
\end{table}

The mean $\mu_{3.5}$, standard deviation $s_{3.5}$ and coefficient of variation $COV = s \slash \vert \mu \vert$ are computed at the end of the loading as the imposed displacement reaches $E=-3.5e^{-3}$. These values are reported in table~\ref{tab:StatResp}. Some noteworthy conclusions can be drawn from these results:
\begin{itemize}
 \item As correlation length $\overline{\ell_c}$ decreases, so does the variability ($COV_{3.5}$) of the macroscopic response.
 \item Case \#3 shows that it is possible to find a set of parameters that satisfies $\overline{\ell_c} \geq \overline{\ell_c}_{min}$ and for which the variability of the macroscopic material response is very small ($COV_{3.5} = 0.6$\%). This means that any E-mesoscale generated in case \#3 yields almost the same material response at macroscale, which can be qualified as a representative response for the boundary conditions considered.
 \item There is a strong reduction of the variability that drops from $COV \geq 50$\% for the material parameters at E-mesoscale to $COV_{3.5} \leq 3.3$\% for the macroscopic material response at maximum compression.
 \item As $\overline{L^0}$ is kept constant and equal to~1 while $\overline{\ell_c}$ decreases (scenario \#1), that are cases \#1, \#2 and \#3 (left column in figure~\ref{fig:ConLaw-1}), mean response changes. On the contrary, as the $\overline{\ell_c} \slash \overline{L^0}$ ratio is kept constant while $\overline{\ell_c}$ decreases (scenario \#2), that are cases \#5, \#4 and \#3 (right column in figure~\ref{fig:ConLaw-1}), mean response remains almost unchanged.
 \item Also, for scenario \#1, variability ($COV_{3.5}$) is less than for scenario \#2 for a same correlation length.
\end{itemize}
Considering same correlation length in scenarios \#1 and \#2, there are still two major differences between both scenarios. Firstly, the discretization of the power spectral density function (equation~\eqref{eq:S0-NA}) is not the same because $\Delta\overline{\kappa} = \overline{\kappa}_u \slash N$ depends on $M$ (see equation~\eqref{eq:ku2}). Secondly, meso-structures have (asymptotically) ergodic properties in mean and correlation for scenario \#1 ($\overline{L^0} = 1$), while this is no more the case in scenario \#2 ($\overline{L^0} \geq 1$).

\subsubsection{Emergence of a macroscopic response}
\label{sec:EmeComRes}

\begin{figure}[htb]
 \begin{tabular}{cc}
  \includegraphics[width=.5\textwidth]{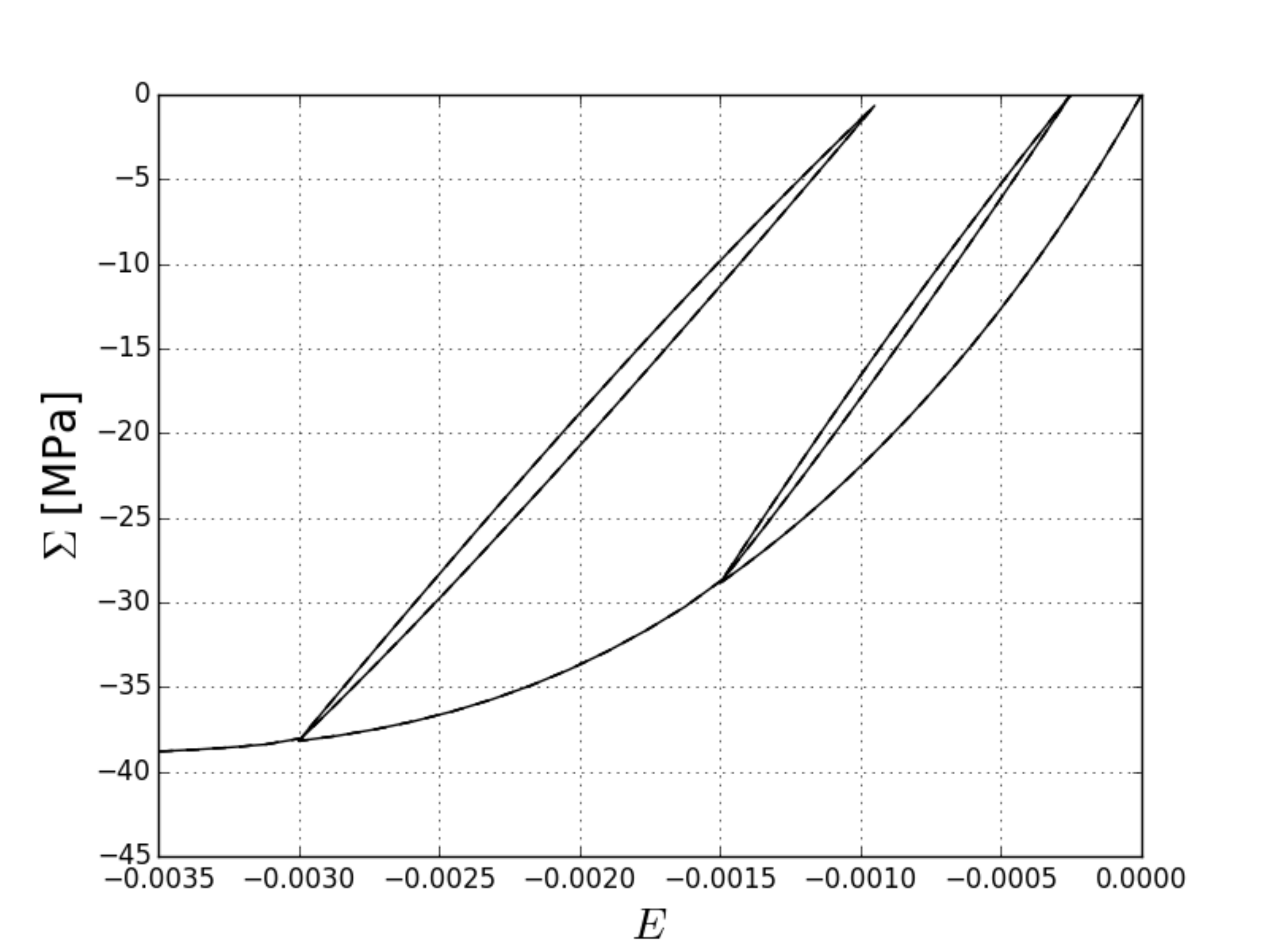} & \includegraphics[width=.5\textwidth]{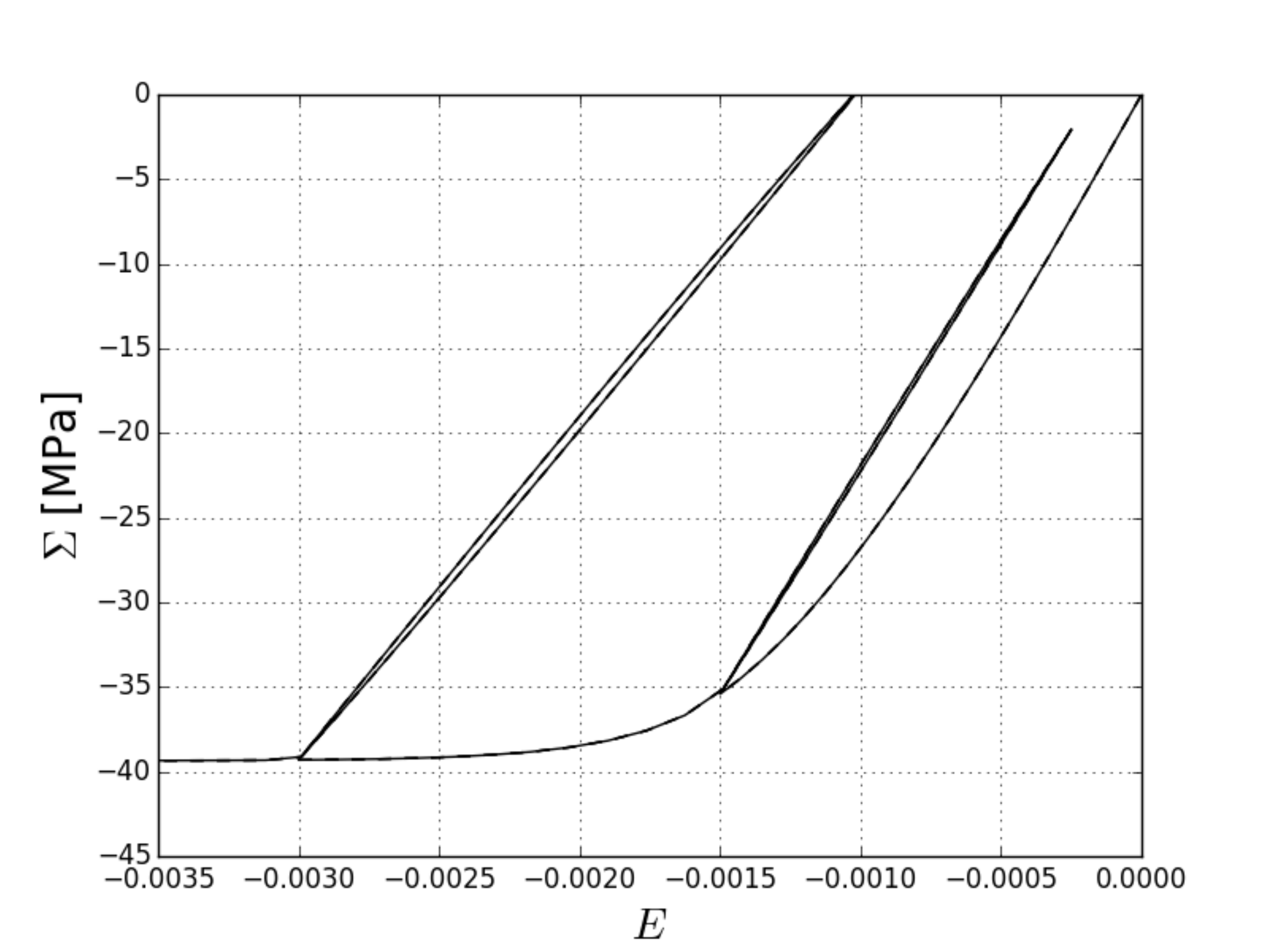}
 \end{tabular}
 \caption{1D macroscopic compressive cyclic response of a concrete elementary area for two different distributions for elastic modulus $C$: [left] log-normal distribution and [right] uniform distribution.}
 \label{fig:CompMacrResp}
\end{figure}

1D macroscopic compressive cyclic response of a concrete elementary area generated with parameters $M=32$ and $\overline{\ell_c}=0.1$ (case \#3) is shown in figure~\ref{fig:CompMacrResp}. Two different distributions for elastic modulus $C$ are considered: i) log-normal distribution as introduced in table~\ref{tab:DisLaw} and ii) uniform distribution $C \sim \mathcal{U}(10e^3, 50e^3)$. Because concrete macroscopic response is more realistic for the log-normal distribution (compare with figure~\ref{fig:Ramtani}), this distribution was adopted for the numerical applications previously shown in section~\ref{sec:ModConRep}. 

Figure~\ref{fig:CompMacrResp} shows that salient features of the experimentally observed concrete behavior (figure~\ref{fig:Ramtani}) are represented by the multi-scale stochastic approach presented in this chapter. An initial elastic phase ($E \leq 0.2e^{-3}$) is followed by nonlinear strain hardening; stiffness degradation is observed when unloading (damage); residual deformation remains after complete unloading (plasticity). Besides, in unloading-reloading cycles, hysteresic behavior is produced. It is interesting to observe that nonlinear hardening along with hysteresis in unloading-reloading cycles at macroscale are not explicitly modeled at E-mesoscale (see figure~\ref{fig:pnt-law}): they emerge from numerous nonlinear and uncertain responses at E-mesoscale consequently to both spatial variability and averaging of the responses at E-mesoscale over $\mR$.

One interesting feature in the macroscopic 1D response of concrete in uniaxial compressive cyclic loading is the hysteresis observed in unloading-reloading cycles: while reloading, the $\Sigma$-$E$ curve follows another path than while unloading, which generates energy dissipation at the structural level. As it is a source of damping in reinforced concrete structures in seismic loading, which modeling is a challenging issue, modeling this hysteresis has been the focus of research work (see e.g. \cite{RagLabMaz2000, Jeh-et-al2010}). In~\cite{JehCot2015}, a simplified version of the stochastic multi-scale material model presented in this chapter has been developed with only the yield stress $\sigma_y$ being heterogenous and without damage-plasticity coupling. The material model has been implemented in a beam element and the capacity of the concrete behavior law to generate structural damping has been shown in the numerical testing of reinforced concrete columns in free vibration. Besides, this has shown that the proposed material model can be used in solving numerical nonlinear dynamic analyses of structural frame elements.

\section{Conclusion}

A stochastic multi-scale approach has been presented in this chapter for numerical modeling of complex materials, that are materials for which macroscopic response results from the interaction of numerous intertwined nonlinear and uncertain mechanisms at lower scales. This approach is based on the construction of an equivalent mesoscale (E-mesoscale) where material properties are heterogenous and where local behavior is nonlinear, coupling mechanisms such as plasticity and damage. Macroscopic response is then computed using averaging formula over an elementary domain (ED). The approach is used to model the uni-dimensional response of concrete material in uniaxial compressive cyclic loading. It is shown that a random E-mesoscale can be generated by spectral representation in such a way that the macroscopic response does not depend on the realization of the random meso-structure. The ED, equipped with such an E-mesoscale, can then be considered as a representative material domain because homogeneous macroscopic properties are retrieved. Besides, this also means that this homogeneous macroscopic behavior is obtained without stochastic homogenization. Because only homogeneous displacements are considered for the boundary conditions for the ED, note that the term ``representative" does not imply here independence of the boundary conditions. Moreover, the macroscopic concrete response modeled by this approach exhibits most of the salient features observed in experimental uniaxial cyclic compressive tests on concrete specimens, and particularly the hysteresis loops observed in unloading-reloading cycles. Considering that some of these features are not explicitly represented at the E-mesoscale, this shows the capacity of the approach for letting macroscopic behaviors emerge from simpler mechanisms at lower scales.

In this chapter, the E-mesoscale for concrete material is built on a conjectural basis. Nevertheless, the assumptions that are made both about the mechanical behavior at this scale and the description of the heterogeneity in the properties yield a macroscopic response that reproduces salient features that can be observed experimentally testing concrete specimen. Consequently, although the proposed approach needs to be fed by experimental evidence, it certainly can also trigger experimental research because it provides a rational explanation of macroscopic mechanisms from lower-scale information.

\section*{acknowledgement}
The author thanks Prof. George Deodatis for fruitful discussions about the content of this book chapter and for hosting him during its preparation. This research is supported by a Marie Curie International Outgoing Fellowship within the 7th European Community Framework Programme (proposal No. 275928).

\section*{Appendix 1: On the ergodicity in correlation of the random fields simulated with equation~\eqref{eq:gVFcomp}}
\addcontentsline{toc}{section}{Appendix 1}
From equation~\eqref{eq:gVFcomp}, $j \in [1,...,m]$:
\begin{align} \label{eq:gVFcomp2}
 g_j(\bx; \theta) = 2 \sqrt{\Delta\kappa_1 \Delta\kappa_2} & \sum_{l=1}^m \sum_{\alpha=1}^2 \sum_{n_1=0}^{N_1-1} \sum_{n_2=0}^{N_2-1} \vert H_{jl}(\bkappa_{n_1n_2}^{\alpha}) \vert \times \nonumber \\
 & \qquad\qquad \cos(\bkappa_{n_1n_2}^{\alpha} \cdot \bx - \varphi_{jl}(\bkappa_{n_1n_2}^{\alpha}) + \Phi_{l, n_1n_2}^{\alpha}(\theta))
\end{align}
On the one hand, because the random phases $\Phi(\theta)$ are independent and uniformly distributed over $[0 \, , \, 2\pi]$, the ensemble correlation function of two sample functions $g_j(\bxi;\theta)$ and $g_k(\bxi;\theta)$ reads:
\begin{align}
R_{jk}(\bxi) &:= \mathbb{E}\left[ g_j(\bx; \theta) \ g_k(\bx + \bxi; \theta) \right] \nonumber \\
& = \frac{1}{4\pi^2} \int_0^{2\pi} \int_0^{2\pi} g_j(\bx ; \theta) \ g_k(\bx+\bxi ; \theta) \ d\Phi^a \, d\Phi^b
\end{align}
On the other hand, the spatial correlation of the two sample functions $g_j(\bxi;\theta)$ and $g_k(\bxi;\theta)$ over a 2D area of size $L^0_1 \times L^0_2$ reads:
\begin{align}
\tilde{R}_{jk}(\bxi) &:= \langle g_j(\bx; \theta) \ g_k(\bx + \bxi; \theta) \rangle_{L^0_1 \times L^0_2} \nonumber \\
 & := \frac{1}{L^0_1 \ L^0_2} \int_0^{L^0_1} \int_0^{L^0_2} g_j(\bx ; \theta) \ g_k(\bx+\bxi ; \theta) \ dx_1 \, dx_2
\end{align}

Then, we have from equation~\eqref{eq:gVFcomp2}:
\begin{align}
 & g_j(\bx; \theta) \ g_k(\bx + \bxi; \theta) = 4 \Delta\kappa_1 \Delta\kappa_2  \nonumber \\
 & \qquad\qquad\qquad\qquad \times \sum_{l^a=1}^m \sum_{\alpha^a=1}^2 \sum_{n_1^a=0}^{N_1-1} \sum_{n_2^a=0}^{N_2-1} \sum_{l^b=1}^m \sum_{\alpha^b=1}^2 \sum_{n_1^b=0}^{N_1-1} \sum_{n_2^b=0}^{N_2-1} \vert H_{jl^a}(\bkappa_{n_1^a n_2^a}^{\alpha^a}) \vert \nonumber \\ 
& \qquad\qquad\qquad\qquad \times \vert  H_{kl^b}(\bkappa_{n_1^b n_2^b}^{\alpha^b}) \vert \ A_{jkl^al^b}(\bkappa_{n_1^a n_2^a}^{\alpha^a}, \bkappa_{n_1^b n_2^b}^{\alpha^b}; \bx, \bxi; \theta)
\end{align}
where we introduced
\begin{align}
 & A_{jkl^al^b}(\bkappa_{n_1^a n_2^a}^{\alpha^a}, \bkappa_{n_1^b n_2^b}^{\alpha^b}; \bx, \bxi; \theta) =  \cos(\bkappa_{n_1^a n_2^a}^{\alpha^a} \cdot \bx - \varphi_{jl^a}(\bkappa_{n_1^a n_2^a}^{\alpha^a}) + \Phi_{l^a, n_1^a n_2^a}^{\alpha^a}(\theta)) \nonumber \\
 & \qquad\qquad \times \cos(\bkappa_{n_1^b n_2^b}^{\alpha^b} \cdot (\bx + \bxi) - \varphi_{kl^b}(\bkappa_{n_1^b n_2^b}^{\alpha^b}) + \Phi_{l^b, n_1^b n_2^b}^{\alpha^b}(\theta)) 
\end{align}
Using now the relation $\cos \beta \, \cos \gamma = \frac{1}{2} \left\{ \cos(\beta + \gamma) + \cos(\beta - \gamma) \right\}$, it comes:
\begin{align}
 & A_{jkl^al^b}(\bkappa_{n_1^a n_2^a}^{\alpha^a}, \bkappa_{n_1^b n_2^b}^{\alpha^b}; \bx, \bxi; \theta) = \frac{1}{2} \Big\{ \cos\left( (\bkappa_{n_1^b n_2^b}^{\alpha^b} + \bkappa_{n_1^a n_2^a}^{\alpha^a}) \cdot \bx + \bkappa_{n_1^b n_2^b}^{\alpha^b} \cdot \bxi - \varphi_{jl^b}(\bkappa_{n_1^b n_2^b}^{\alpha^b}) \right. \nonumber \\ 
 & \qquad \left. - \varphi_{kl^a}(\bkappa_{n_1^a n_2^a}^{\alpha^a}) + \Phi_{l^b, n_1^b n_2^b}^{\alpha^b}(\theta) + \Phi_{l^a, n_1^a n_2^a}^{\alpha^a}(\theta) \right) + \cos\left( (\bkappa_{n_1^b n_2^b}^{\alpha^b} - \bkappa_{n_1^a n_2^a}^{\alpha^a}) \cdot \bx \right.\nonumber \\
 & \qquad \left. + \bkappa_{n_1^b n_2^b}^{\alpha^b} \cdot \bxi - \varphi_{jl^b}(\bkappa_{n_1^b n_2^b}^{\alpha^b}) + \varphi_{kl^a}(\bkappa_{n_1^a n_2^a}^{\alpha^a}) + \Phi_{l^b, n_1^b n_2^b}^{\alpha^b}(\theta) - \Phi_{l^a, n_1^a n_2^a}^{\alpha^a}(\theta) \right) \Big\}
\end{align}

To calculate the ensemble correlations $R_{jk}(\bxi)$, we have to calculate:
\begin{equation}
 B_{jkl^al^b}(\bkappa_{n_1^a n_2^a}^{\alpha^a}, \bkappa_{n_1^b n_2^b}^{\alpha^b}; \bx, \bxi) = \int_0^{2\pi} \int_0^{2\pi} A_{jkl^al^b}(\bkappa_{n_1^a n_2^a}^{\alpha^a}, \bkappa_{n_1^b n_2^b}^{\alpha^b}; \bx, \bxi; \theta) \ d\Phi^a \, d\Phi^b
\end{equation}
Because functions $A_{jkl^al^b}$ are periodic of period $2\pi$, functions $B_{jkl^al^b} = 0$ except in the case where $\Phi_{l^b, n_1^b n_2^b}^{\alpha^b}(\theta) = \Phi_{l^a, n_1^a n_2^a}^{\alpha^a}(\theta)$, that is as $n_1^a = n_1^b = n_1$ and $n_2^a = n_2^b = n_2$ and $\alpha^a = \alpha^b = \alpha$ and $l^a = l^b = l$. This yields:
\begin{equation}
 B_{jkl^al^b}(\bkappa_{n_1^a n_2^a}^{\alpha^a}, \bkappa_{n_1^b n_2^b}^{\alpha^b}; \bx, \bxi) = 2 \pi^2 \cos\left( \bkappa_{n_1 n_2}^{\alpha} \cdot \bxi - \varphi_{jl}(\bkappa_{n_1 n_2}^{\alpha}) + \varphi_{kl}(\bkappa_{n_1 n_2}^{\alpha}) \right)
\end{equation}
and, finally:
\begin{equation}
 R_{jk}(\bxi) = 2 \Delta\kappa_1 \Delta\kappa_2 \sum_{l=1}^m \sum_{\alpha=1}^2 \sum_{n_1=0}^{N_1-1} \sum_{n_2=0}^{N_2-1} \cos\left( \bkappa_{n_1 n_2}^{\alpha} \cdot \bxi - \varphi_{jl}(\bkappa_{n_1 n_2}^{\alpha}) + \varphi_{kl}(\bkappa_{n_1 n_2}^{\alpha}) \right)
\end{equation}

Then, to calculate the spatial correlations $\tilde{R}_{jk}(\bxi)$, we have to calculate:
\begin{equation}
 \tilde{B}_{jkl^al^b}(\bkappa_{n_1^a n_2^a}^{\alpha^a}, \bkappa_{n_1^b n_2^b}^{\alpha^b}; \bxi; \theta) = \int_0^{L^0_1} \int_0^{L^0_2} A_{jkl^al^b}(\bkappa_{n_1^a n_2^a}^{\alpha^a}, \bkappa_{n_1^b n_2^b}^{\alpha^b}; \bx, \bxi; \theta) \ dx_1 \, dx_2
\end{equation}
Because functions $A_{jkl^al^b}$ are periodic of period $L^0_1 \times L^0_2$, and with the condition that $H_{jk}=0$ as any $\kappa_i = 0$, for any $(j,k) \in [1,...,m]^2$ and for any $i \in [1,...,d]$, functions $\tilde{B}_{jkl^al^b}$ are equal to zero, except if $\bkappa_{n_1^b n_2^b}^{\alpha^b} = \bkappa_{n_1^a n_2^a}^{\alpha^a}$, that is $n_1^a = n_1^b = n_1$ and $n_2^a = n_2^b = n_2$ and $\alpha^a = \alpha^b = \alpha$, in which case:
\begin{align} \label{eq:Btilde}
 \tilde{B}_{jkl^al^b}(\bkappa_{n_1^a n_2^a}^{\alpha^a}, \bkappa_{n_1^b n_2^b}^{\alpha^b}; \bxi; \theta) &= \frac{L^0_1 \ L^0_2}{2} \cos\left( \bkappa_{n_1 n_2}^{\alpha} \cdot \bxi - \varphi_{jl^b}(\bkappa_{n_1 n_2}^{\alpha}) \right. \nonumber \\
 & \qquad \left.+ \varphi_{kl^a}(\bkappa_{n_1 n_2}^{\alpha}) + \Phi_{l^b, n_1 n_2}^{\alpha}(\theta) - \Phi_{l^a, n_1 n_2}^{\alpha}(\theta) \right)
\end{align}

With this expression of the functions $\tilde{B}_{jkl^al^b}$, we do not have $\tilde{R}(\bxi) = R(\bxi)$. However, when wave-number shifts are introduced so that wave numbers $\bkappa$ become dependent on the index $l$ (as in~\cite{Deodatis1996b, PopDeoPre1998}), the condition $l^a = l^b = l$ has to be added for $\tilde{B}_{jkl^al^b}$ not to be equal to zero. Consequently, $\Phi_{l^b, n_1 n_2}^{\alpha}(\theta) = \Phi_{l^a, n_1 n_2}^{\alpha}(\theta)$ in equation~\eqref{eq:Btilde} and we finally recover $R(\bxi) = \tilde{R}(\bxi)$, meaning that sample fields $g_j(\bxi; \theta)$ are ergodic in correlation.

\section*{Appendix 2: Material model at mesoscale for the numerical applications}
\addcontentsline{toc}{section}{Appendix 2}
For the one-dimensional material model at mesoscale used in the numerical applications shown in section~\ref{sec:NumApp}, we use $h(\sigma)=\vert \sigma \vert$ in the definition of the criterium function (see equation~\eqref{eq:cri-fun} in section~\ref{sec:MatModMes}):
\begin{equation}\label{eq:Phi-1D}
 \phi_{n+1} = \vert \sigma_{n+1} \vert - \sigma_y \quad \Rightarrow \quad \nu_{n+1} := \frac{\partial \phi_{n+1}}{\partial \sigma_{n+1}} = sign(\sigma_{n+1})
\end{equation}
Then, equation~\eqref{eq:sig} can be written as:
\begin{equation}\label{eq:SigSigTr}
 \sigma_{n+1} = \sigma^{trial}_{n+1} - D^{-1}_n \gamma_{n+1} sign(\sigma_{n+1})
\end{equation}
Multiplying both sides of equation~\eqref{eq:SigSigTr} by $sign(\sigma_{n+1})$, it comes:
\begin{equation}\label{eq:SignSign}
 \vert \sigma_{n+1} \vert = \sigma^{trial}_{n+1} sign(\sigma_{n+1}) - D^{-1}_n \gamma_{n+1}
\end{equation}
Multiplying now both sides of equation~\eqref{eq:SignSign} by $sign(\sigma^{trial}_{n+1})$, it comes:
\begin{equation}
 \left( \vert \sigma_{n+1} \vert + D^{-1}_n \gamma_{n+1} \right) sign(\sigma^{trial}_{n+1}) = \vert \sigma^{trial}_{n+1} \vert sign(\sigma_{n+1})
\end{equation}
Setting $\gamma_0 = 0$ and $D_0 > 0$, $\left( \vert \sigma_{n+1} \vert + D^{-1}_n \gamma_{n+1} \right)$ necessarily is non-negative because $\dot{\gamma} \geq 0$ and $\dot{D} \vert \sigma \vert = r \, \dot{\gamma} \geq 0$. Consequently:
\begin{equation}\label{eq:sign}
 sign(\sigma_{n+1}) = sign(\sigma^{trial}_{n+1})
\end{equation}
Then, we have from equations~\eqref{eq:SigSigTr} and~\eqref{eq:Phi-1D}:
\begin{align}
 & \vert \sigma_{n+1} \vert = \vert \sigma^{trial}_{n+1} \vert - D^{-1}_n \gamma_{n+1} \\
 & \phi_{n+1} = \phi^{trial}_{n+1} - D^{-1}_n \gamma_{n+1}
\end{align}
with $\phi^{trial}_{n+1} = \vert \sigma^{trial}_{n+1} \vert - \sigma_y$, from which we can calculate $\gamma_{n+1}$ in case of inelastic evolution:
\begin{equation}\label{eq:gam}
 \phi_{n+1} = 0 \quad \Rightarrow \quad \gamma_{n+1} = D_n \, \phi^{trial}_{n+1}
\end{equation}

\section*{Appendix 3: Translation from Gaussian to uniform distributions}
\addcontentsline{toc}{section}{Appendix 3}
Let $\mathfrak{a}_1$ and $\mathfrak{a}_2$ be two independent normal Gaussian variables: $(\mathfrak{a}_1, \mathfrak{a}_2) \sim \mN(0,1)$. Then $\mathfrak{b} = \exp(-(\mathfrak{a}_1^2 + \mathfrak{a}_2^2) \slash 2)$ is a random variable with uniform distribution in $[0, \, 1]$: $\mathfrak{b} \sim \mU(0,1)$. Indeed:
\begin{equation} \label{eq:Pro-b}
 \Pr[\mathfrak{b} \leq b] = \frac{1}{2\pi} \int_{ \{ (a_1, a_2) \vert e^{-\frac{1}{2}(a_1^2 + a_2^2)} \leq b \} } e^{-\frac{1}{2}(a_1^2 + a_2^2)} da_1 da_2 \\
\end{equation}
Then:
\begin{itemize}
 \item if $b > 1$, $\Pr[\mathfrak{b} \leq b] = 1$ because $e^{-\frac{1}{2}(a_1^2 + a_2^2)} \leq 1$, $\forall (a_1, a_2) \in \mathbb{R}^2$;
 \item if $b < 0$, $\Pr[\mathfrak{b} \leq b] = 0$ because $e^{-\frac{1}{2}(a_1^2 + a_2^2)} > 1$, $\forall (a_1, a_2) \in \mathbb{R}^2$;
 \item and, if $0\leq b \leq 1$, we can rewrite relation~\eqref{eq:Pro-b} with polar coordinates as:
\end{itemize} 
\begin{equation}
 \Pr[\mathfrak{b} \leq b] = \frac{1}{2\pi} \int_0^{2\pi} \int_{\sqrt{-2 \ln b}}^{+\infty} \ e^{-\frac{1}{2} r^2} \ r \, dr d\theta = \left[-e^{-\frac{1}{2} r^2}\right]_{\sqrt{-2 \ln b}}^{+\infty} = b \nonumber
\end{equation}

\end{document}